\documentclass[aps, prd, twocolumn, showpacs, floatfix, letterpaper,
 nofootinbib, amsmath, amssymb, superscriptaddress]{revtex4-2}
\usepackage{amssymb,amsmath}
\usepackage{latexsym}
\usepackage{graphicx}
\usepackage{caption}
\usepackage{subcaption}
\usepackage{enumerate}
\usepackage{float}

\usepackage{xcolor}
\usepackage{graphics}
\usepackage{epsfig}
\usepackage{bm}  
\usepackage{ulem}

\begin{document}

\newcommand{\ud}{\mathrm{d}}

\title{Dynamics of primordial fields in quantum cosmological spacetimes}

\author{Przemys\l aw Ma\l kiewicz}
\email{Przemyslaw.Malkiewicz@ncbj.gov.pl}
 \author{Artur Miroszewski}
 \email{Artur.Miroszewski@ncbj.gov.pl}
\affiliation{National Centre for Nuclear Research, Pasteura 7, 02-093 Warsaw, Poland} 

\date{\today}

\begin{abstract}
Quantum cosmological models are commonly described by means of semiclassical approximations in which a smooth evolution of the expectation values of elementary geometry operators replaces the classical and singular dynamics. The advantage of such descriptions is that they are relatively simple and display the classical behavior for large universes. However, they may smooth out an important inner structure and to include it a more detailed treatment is needed. The purpose of the present work is to investigate { quantum uncertainty in the basic background variables} and its influence on primordial gravitational waves. To this end we quantize a model of the Friedmann-Lemaitre-Robertson-Walker universe filled with a linear barotropic cosmological fluid and with gravitational waves.  { We carefully derive the dynamical equations for the perturbations in quantum spacetime.}  The quantization yields an equation of motion for the Fourier modes of gravitational radiation, which is a quantum extension to the usual parametric oscillator equation for gravitational waves propagating in an expanding universe. The two quantum effects from the cosmological background that enter the enhanced equation of motion are (i) a repulsive potential resolving the big bang singularity and replacing it with a big bounce and (ii) uncertainties in the numerical values for the background spacetime dynamical variables. First we study the former effect and its consequences for the primordial amplitude spectrum and carefully discuss the relation between the bounce scale and the physical predictions of the model. Next we investigate the latter effect, in particular the extent to which it may affect the primordial amplitude of gravitational waves. Making use of the WKB approximation we find an analytical formula for the amplitude spectrum as a function of the quantum dispersion of the background spacetime.\end{abstract}


\maketitle

\section{Introduction}
Theories of the origin of primordial structure that are based on models of a quantum bounce replacing the big bang singularity (see e.g.  \cite{Peter2008,Ashtekar2012}) are often formulated in terms of the ``effective" or ``trajectory" dynamics of early Universe. The goal of the present work is to construct and study an enhanced framework that incorporates a full quantum description of the homogenous cosmological spacetime and its full action on the perturbations to homogeneity propagating thereon (see e.g. \cite{Peter2005,Ashtekar:2009mb,Brizuela:2015tzl,Martinez:2016hmn} for other proposals).\\

\begin{figure}[t]
\centering
\includegraphics[width=0.47\textwidth]{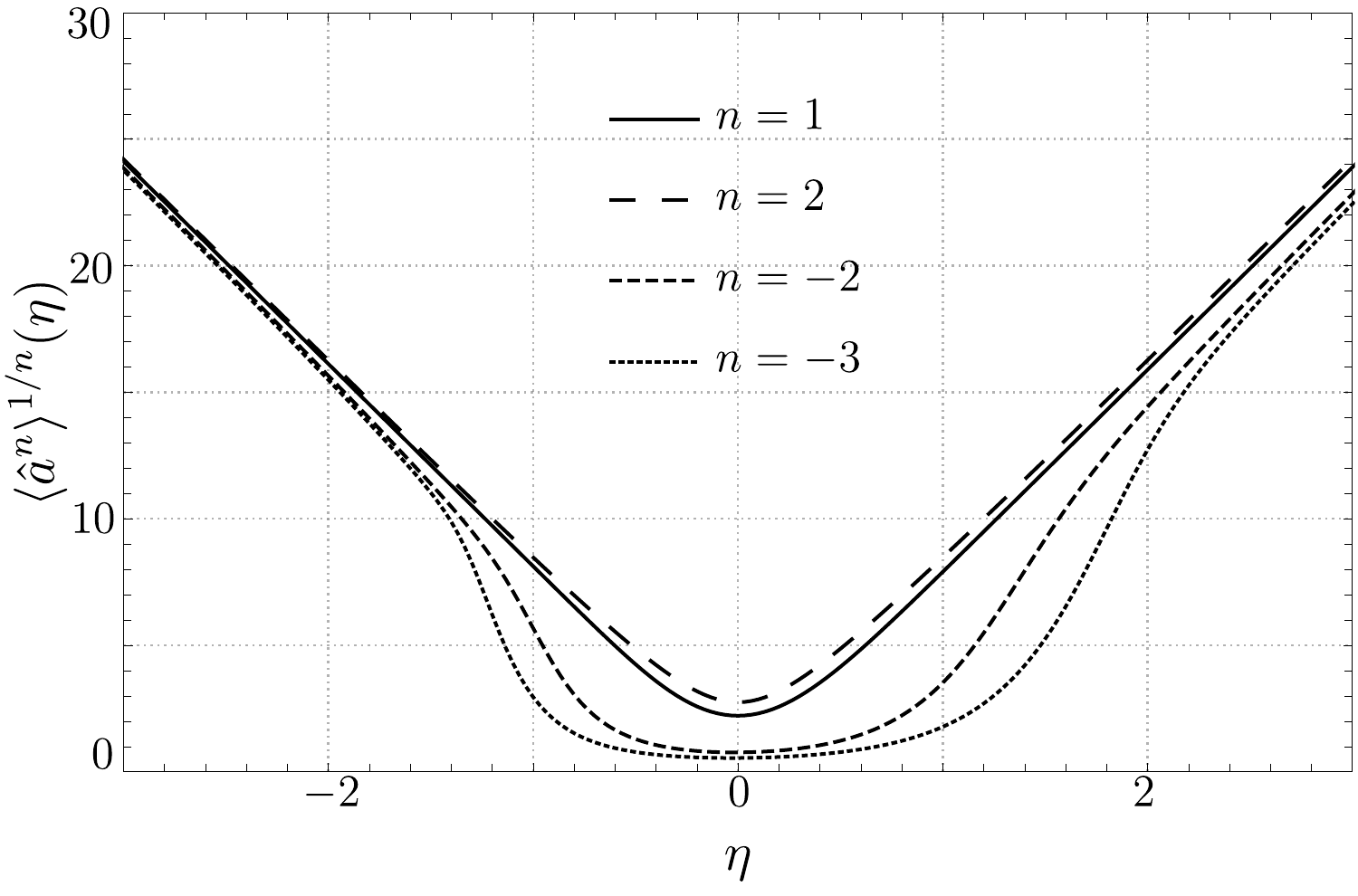}
\caption{semiclassical dynamics of the scale factor obtained from various quantum dynamical variables of which all satisfy the classical limit for large volumes. (All the plots were obtained for a unique wave function of Eq. (\ref{full_solution}) with $K=\frac{3}{4}$, $\sigma=2$,  $x_0=30$, $p_0=-4$.)}
\label{fig:intro}
\end{figure}

There are two distinct consequences of the description of the background spacetime by means of a wave function. First, the singular dynamics of elementary classical variables is replaced with nonsingular dynamics of quantum expectation values yielding semiclassical bouncing trajectories. This aspect of quantum cosmological spacetimes and its effect on the propagation of quantum fields has been widely studied for cosmological applications. Second, the background spacetime wave function implies some spread in the background dynamical variables and, in particular, in the coupling between the perturbations and the background mode. The consequences of the latter are rarely studied \cite{Gomar:2016rso,Ashtekar2016}. We will illustrate the origin of this effect with a simple example. Note that there are many ways in which the classical cosmological evolution in terms of the scale factor $a$ may be replaced by a semiclassical evolution of $a$. For instance, the classical scale factor may be replaced with the expectation values of various powers of the quantum scale factor as follows: $a(\eta)=\langle \widehat{a}^n(\eta)\rangle^{1/n}$, where $n$ is a nonzero value. In Fig. \ref{fig:intro} we plot the evolution of the scale factor in conformal time for a unique wave function and a few values of $n$. The plot shows, in particular, that for negative values of $n$, the Universe generically undergoes a phase of accelerated contraction before being decelerated, halted and pushed into expansion, and that the dynamics may exhibit a degree of asymmetry between its contracting and expanding phases. This is a purely quantum spread effect which demonstrates that ``quantum forces" are not necessarily purely repulsive even when they ultimately revert the dynamics of the Universe. The ambiguity illustrated by this example is neglected by semiclassical trajectories in which all the above scale factors evolve the same. It is therefore necessary to find if this neglected structure could produce some observable cosmological effects. \\

In this work we consider cosmological implications of the presence of quantum uncertainties in a universe undergoing a bounce. We omit the nonessential, though possible, phase of inflation and instead focus on fluid-dominated universes. Moreover, we restrict our attention to the universe from which the density perturbations are absent. In other words, we investigate the effect of quantum uncertainties of the background spacetime on the dynamical law of primordial gravitational waves in fluid-driven bouncing universes. Quantum bounces in such universes have been previously studied within the Bohm-de Broglie approach in \cite{Peter2006}. The results obtained therein are in fact reproduced in a semiclassical limit of our model. We go beyond the semiclassical description and add spread to the background which produces an extra structure in the dynamical coupling between the gravitational waves and the background. As we shall see, it influences the evolution of the amplitude of primordial gravitational waves and their final state. It is clear that the existence of this influence must be universal to all quantum cosmological models irrespective of the employed quantization procedure or the assumed background symmetries.\\

The outline of the paper is as follows. In Sec. II we briefly describe the Hamiltonian formalism for the investigated cosmological model and its quantization. Our discussion includes the issue of backreaction and entanglement between the background spacetime and the perturbations. We also discuss the existence of the classical limit which is necessary for cosmological applications. The main result of this section is the quantum evolution equation for the modes of gravitational radiation. In Sec. III we first employ a semiclassical method based on infinitesimally narrow wave packets to study the quantum bounce and the resultant quantum evolution equation. We numerically solve that equation and discuss the cosmological implications of the obtained result. Then we employ the full quantum approach and discuss the new qualitative features that it brings in at the level of the aforementioned equation. We resort to the WKB approximation in order to analytically investigate the evolution of the gravity-wave amplitude in a function of the spread of the quantum background. The main findings are summarized and discussed in Sec. IV.

\section{Quantum cosmological model}
 
 \subsection{Classical and quantum Hamiltonian}
 
Let us assume a flat universe with toroidal topology $\Sigma=\mathbb{T}^3$ and the line element
\begin{align}
\ud s^2=-N^2\ud t^2+a^2(\delta_{ab}+h_{ab}(x))\ud x^a\ud x^b,
\end{align} 
where the coordinate volume equals $\int_{\Sigma}\ud^3x=\mathcal{V}_0$ and the physical volume equals $V=a^3\mathcal{V}_0$. The metric perturbations $h_{ab}$ and their conjugate momenta ${\pi}^{ab}$ are resolved into the Fourier coefficients
\begin{align}\begin{split}
\check{h}_{ab}(\vec{k})=\mathcal{V}_0^{-1}\int_{\Sigma} {h}_{ab}(x)e^{-i\vec{k}\vec{x}}\ud^3x,\\
\check{\pi}^{ab}(\vec{k})=\int_{\Sigma} {\pi}^{ab}(x)e^{-i\vec{k}\vec{x}}\ud^3x,\end{split}
\end{align}
which are next expressed in a new tensorial basis with two distinct polarization modes of the gravitational wave,
\begin{align}
 \check{h}_{\pm}=\check{h}_{ab}A^{ab}_{\pm},~~~\check{\pi}_{\pm}=\check{\pi}^{ab}A_{ab}^{\pm},
 \end{align}
where $A^{ab}_{+}=\frac{1}{\sqrt{2}}(v^a w^b+v^b w^a)$, $A^{ab}_{-}=\frac{1}{\sqrt{2}}(v^a v^b-w^b w^a)$, and $\vec{v}$ and $\vec{w}$ are such that $|\vec{k}|^{-1}\vec{k}$, $\vec{v}$ and $\vec{w}$ form an orthonormal frame with respect to the fiducial metric $\delta_{ab}$. The new variables satisfy the usual commutation relation, $\{\check{\pi}_{\pm}(\vec{k}), \check{h}_{\pm}(\vec{l})\}=\delta_{\pm,\pm}\cdot\delta_{\vec{k},-\vec{l}}$, and the reality condition for the field ${h}_{ab}(x)$ implies $\check{h}_{\pm}(\vec{k})=\check{h}_{\pm}^*(-\vec{k})$ and $\check{\pi}_{\pm}(\vec{k})=\check{\pi}_{\pm}^*(-\vec{k})$.\\

The physical Hamiltonian for the fluid-driven homogeneous and isotropic universe with linear tensor perturbations thereon reads \cite{Bergeron:2017ddo},
\begin{align}\label{ham}
\mathbf{H}=\mathbf{H}^{(0)}+\sum_{\vec{k}}\mathbf{H}^{(2)}_{\vec{k}},
\end{align}
where
\begin{align}\label{defs}\begin{split}
\mathbf{H}^{(0)}&=\mathfrak{g} p^2,\\
\mathbf{H}^{(2)}_{\vec{k}}&=-\mathfrak{g}\left(\frac{q}{\gamma}\right)^{-2}|\check{\pi}_{\pm}(\vec{k})|^2-\frac{k^2}{4\mathfrak{g}}\left(\frac{q}{\gamma}\right)^{\frac{6w+2}{3-3w}}|\check{h}_{\pm}(\vec{k})|^2,\end{split}
\end{align}
where $\mathfrak{g}=\frac{16\pi G}{\mathcal{V}_0}$, and $w$ is the ratio of the fluid's pressure to its energy density $\gamma=\frac{4\sqrt{6}}{3(1-w)}$, whereas $q=\gamma a^{\frac{3-3w}{2}}$ and $p=\frac{3(1-w)\gamma}{8\mathfrak{g}} a^{\frac{3+3w}{2}}\textrm{H}$ (where $\textrm{H}=\frac{\dot{a}}{Na}$ is the Hubble rate) are canonical background variables. It follows from the Friedmann equation that the background Hamiltonian $\mathbf{H}^{(0)}=\frac{a^{3+3w}}{16\mathfrak{g}}\textrm{H}^2=\frac{a^{3+3w}}{96}\rho\mathcal{V}_0$ equals $1/96$ of the energy of matter in the entire universe when its physical and coordinate volumes are equal, $V=\mathcal{V}_0$. The Hamiltonian (\ref{ham}) generates the dynamics with respect to a fluid variable that has been removed from the phase space. This choice of internal clock variable yields the lapse $N=a^{3w}$.\\

We fix the coordinates by setting $\mathcal{V}_0=l_P^3$, i.e., the coordinate volume equals the Planck volume. Furthermore, we assume that the present volume of the universe equals $V_0=r\cdot 1.25\cdot 10^{185}l_P^3$, where $r>1$ is the ratio of the volume of the universe to the volume of its observable patch. This implies the present value of the scale factor to be $a_0=5\cdot 10^{61}r^{1/3}$. We set the pivot scale to correspond to a tenth of the diameter of the observable universe, i.e $\lambda_{*,phys}=5\cdot 10^{60}l_P$, which  yields the coordinate pivot wave number $k_{*}=20\pi r^{1/3} l_P^{-1}$. Given the present value of the Hubble rate, $\mathrm{H}=11.5\cdot 10^{-62}l_P^{-1}$, and the redshift of the matter-radiation equality era, $z_{eq}=3400$, we are able to estimate the value of the Hamiltonian for the radiation-dominated universe (i.e. $w=\frac{1}{3}$),
\begin{align}
\mathbf{H}^{(0)}=2.3\cdot 10^{120}r^{\frac{4}{3}}m_P,
\end{align}
where $r$ needs still to be determined. It follows that if the radiation-dominated era in the expanding universe begins at the volume ${V}_{T}$ with a transition from another fluid-dominated era with $w$ then the primordial value of the Hamiltonian must read\footnote{We apply the Israel junction conditions at the transition between different fluid-dominated cosmological spacetimes. { We do not assume any particular mechanism for the transition, we simply consider a single cosmic fluid with an effective equation of state, which at some point of cosmological expansion turns into radiation.}}
\begin{align}
\mathbf{H}^{(0)}_{w}= 2.3\cdot 10^{120}r^{\frac{4}{3}}m_P {V}_{T}^{w-\frac{1}{3}}
\end{align}
(where the dimensionless $V_T$ gives the number of Planck volumes). Although the value of $r$ is irrelevant for the classical dynamics of the model, the quantum corrections that we study below must depend on it as does the value of the canonical variable $q~\propto~ r^{1/3}$. Therefore, quantum cosmological dynamics depends on the size of the entire universe.
\\

The Hamilton equations generated by the classical Hamiltonian (\ref{ham}) yield the following gravitational wave propagation equation in conformal time, $\eta=\int\left(\frac{q}{\gamma}\right)^{\frac{6w-2}{3-3w}}\ud t$,
\begin{align}\label{wave}\boxed{
\mu_{\pm,\vec{k}}''+\left(k^2-\frac{(q^{\frac{2}{3-3w}})''}{q^{\frac{2}{3-3w}}}\right)\mu_{\pm,\vec{k}}=0,}
\end{align}
where ${\mu}_{\pm,k}=\left(\frac{q}{\gamma}\right)^{\frac{2}{3-3w}}{h}_{\pm,k}$. As we show below, introducing quantum effects to the background dynamics changes this equation in a significant way.\\

Quantization of the Hamiltonian (\ref{ham}) may be carried out as follows. The phase space is the Cartesian product of the homogenous and inhomogeneous sector, $(q,p)\times\prod (\check{h}_{\pm,\vec{k}},\check{\pi}_{\pm,\vec{k}})$. Note that the background canonical variables have a nontrivial range, $(q,p)\in\mathbb{R}_+\times\mathbb{R}$. In this case, the canonical prescription that tells us to replace $q$ and $p$ with the usual position and momentum operators, $\widehat{Q}$ and $\widehat{P}$, does not work properly for the following reasons: (i) the momentum operator on the half line is not self-adjoint and thus it cannot be considered as an elementary observable; (ii) the Hamiltonian operator as the square of the momentum operator is not self-adjoint either and requires imposing a suitable boundary condition on the wave functions. It seems more appropriate to use the dilation instead of the momentum operator, $\widehat{D}=\frac{1}{2}(\widehat{Q}\widehat{P}+\widehat{P}\widehat{Q})$. The dilation operator is self-adjoint and the Hamiltonian operator which is the square of the ratio of dilation to position ``$(\frac{\widehat{D}}{\widehat{Q}})^2$", is self-adjoint for a wide class of symmetric orderings. The quantum zero-order Hamiltonian can be shown to generically contain a purely quantum term,
\begin{align}\label{qh0}
p^2\mapsto \widehat{P}^2+\hbar^2\frac{K}{\widehat{Q}^2},~~~K>0,
\end{align}
which is a repulsive potential $\propto~ \widehat{Q}^{-2}$. The new term prevents the universe from reaching the singularity and generically replaces it with a bounce. More details on the above quantization and the unitary dynamics generated by the quantum Hamiltonian (\ref{qh0}) may be found in \cite{Bergeron:2013ika}.

Quantization of the perturbation variables is straightforward as they have the usual ranges which means that the canonical prescription works well in their case. Thus, $\check{h}_{\pm}(\vec{k})$ and $\check{\pi}_{\pm}(-\vec{k})$ are replaced with the usual position and momentum operators on the real line. Finally, the total quantum Hamiltonian reads \cite{Bergeron:2017ddo}
\begin{align}\label{qham}\begin{split}
\mathbf{H}~\mapsto~&\widehat{\mathbf{H}}=\widehat{\mathbf{H}}^{(0)}+\sum_{\vec{k}}\widehat{\mathbf{H}}^{(2)}_{\vec{k}},\\
\widehat{\mathbf{H}}^{(0)}=& ~\mathfrak{g}\left(\widehat{P}^2+\frac{\hbar^2 K}{\widehat{Q}^2}\right),\\
\widehat{\mathbf{H}}^{(2)}_{\vec{k}}=&-\mathfrak{g}\left(\frac{\widehat{Q}}{\gamma}\right)^{-2}|\widehat{\pi}_{\pm}(\vec{k})|^2-\frac{k^2}{4\mathfrak{g}}\left(\frac{\widehat{Q}}{\gamma}\right)^{\frac{6w+2}{3-3w}}|\widehat{h}_{\pm}(\vec{k})|^2.\end{split}
\end{align}
{ The Hilbert space is given by the tensor product $ \mathcal{H}_{hom}\otimes\mathcal{H}_{inhom}$, where $\mathcal{H}_{hom}$ and $\mathcal{H}_{inhom}$ stand for the background and perturbation Hilbert spaces, respectively.  The background Hilbert space $\mathcal{H}_{hom}=\mathcal{L}^2(\mathbb{R}_+,\ud q)$ is given by the square-integrable functions on the half line, $q>0$. The perturbation Hilbert space $\mathcal{H}_{inhom}=\prod_{\vec{k},\pm}\mathcal{L}^2(\mathbb{R},\ud {h}_{\pm}(\vec{k}))$ is the product of the usual Hilbert spaces  given by the square-integrable functions on the real line. Note that the operator $\widehat{\mathbf{H}}^{(0)}$ acts nontrivially on the states of the background geometry and is a $c$ number while acting on the states of the perturbations. On the other hand, the operator $\widehat{\mathbf{H}}^{(2)}_{\vec{k}}$ acts nontrivially on both the background geometry and the perturbations.}

\subsection{Quantum dynamics}\label{dynamics}
The Hamiltonian (\ref{qham}) is valid if the perturbation variables and their spatial derivatives are much smaller than the unity, and their energy satisfies the following relation:
\begin{align}\begin{split}\label{small}
\big|\mathbf{H}^{(2)}\big|\ll \mathbf{H}^{(0)}.\end{split}
\end{align} 
{ This is consistent with the assumption that the backreaction of the perturbations on the background should be neglected. This makes the Hamilton equations, or equivalently Eq. (\ref{wave}), identical with the linearized Einstein equations (see e.g. \cite{Mukhanov:1990me}), and the backreaction is to be deduced from higher-order dynamics\footnote{Recall that the reduced Hamiltonian is obtained by solving the constraints at linear order and assuming that all quadratic and higher-order terms are negligible in the equations of motion. Hence, in order to properly account for the backreaction effect, one needs to go beyond linear order, which includes solving quadratic or higher-order dynamical constraints as well.}. Similarly, we will impose the lack of backreaction at the quantum level. 

Let us for the moment assume that the state is given by the product of a background state and a perturbation state for all times,
\begin{align}\label{product}
 |\psi\rangle = |\psi_B\rangle\cdot|\psi_P\rangle\in\mathcal{H} \subset \mathcal{H}_{hom}\otimes\mathcal{H}_{inhom}.
\end{align}
This assumption breaks the Schr\"odinger equation produced by the quantum Hamiltonian (\ref{qham}). We determine the dynamical law confined to the product states (\ref{product}) by applying the variational method. We introduce the quantum action
\begin{align}
S_Q(\psi_B,\psi_P):=\int\langle \psi_B, \psi_P|~i\hbar\frac{\partial}{\partial t}-\widehat{\mathbf{H}}~|\psi_B,\psi_P\rangle~\ud t,
\end{align}
whose variation leads to the dynamical equations\footnote{ We could as well restrict the variations to $\delta \psi_P$ because the dynamics of the background has been assumed to be independent and thus, generated by $\widehat{\mathbf{H}}^{(0)}$ alone.}
\begin{align}\begin{split}
i\hbar\frac{\partial}{\partial t} |\psi_B\rangle = \widehat{\mathbf{H}}^{(0)}|\psi_B\rangle + \langle\psi_P|\widehat{\mathbf{H}}^{(2)}|\psi_P\rangle\cdot |\psi_B\rangle,~~\\
i\hbar\frac{\partial}{\partial t} |\psi_P\rangle = \langle\psi_B|\widehat{\mathbf{H}}^{(0)}|\psi_B\rangle\cdot |\psi_P\rangle +\langle\psi_B|\widehat{\mathbf{H}}^{(2)}|\psi_B\rangle\cdot |\psi_P\rangle,\end{split}
\end{align}
which may be further simplified,
\begin{subequations}
\begin{align}\label{eom1}
i\hbar\frac{\partial}{\partial t} |\psi_B\rangle& =  \widehat{\mathbf{H}}^{(0)}|\psi_B\rangle,\\ \label{eom11}
i\hbar\frac{\partial}{\partial t} |\psi_P\rangle &= \langle\psi_B|\widehat{\mathbf{H}}^{(2)}|\psi_B\rangle\cdot |\psi_P\rangle,
\end{align}
\end{subequations}}
{if the backreaction term $\langle\psi_P|\widehat{\mathbf{H}}^{(2)}|\psi_P\rangle\cdot |\psi_B\rangle$ is removed, and the term $ \langle\psi_B|\widehat{\mathbf{H}}^{(0)}|\psi_B\rangle\cdot|\psi_P\rangle$ is discarded as it only adds an overall phase factor to the state $|\psi_B\rangle\cdot|\psi_P\rangle$. It follows that the background state $|\psi_B\rangle$ has to be determined solely from the zero-order Hamiltonian in accordance with our initial assumption. Note that the expectation value $\langle\psi_B|\widehat{\mathbf{H}}^{(2)}|\psi_B\rangle$ is an operator only on $\mathcal{H}_{inhom}$. The key difference brought by the quantum framework is that instead of being based on the classical solutions to the background geometry the perturbation Hamiltonian $\langle\psi_B|\widehat{\mathbf{H}}^{(2)}|\psi_B\rangle$ involves now the expectation values of the background dynamical variables.

The simple product state (\ref{product}) does not exhaust all the possible states in the framework and, in general, will evolve into an entangled state. In fact, the most general state that may satisfy the assumption of the lack of backreaction reads
\begin{align}\label{genproduct}
|\psi_B^{(1)}\rangle\cdot|\psi_P^{(1)}\rangle+|\psi_B^{(2)}\rangle\cdot|\psi_P^{(2)}\rangle+\dots,
\end{align}
and the application of the variational method leads to the following equations:
\begin{widetext}
\begin{align}\begin{split}\label{matrixlaw}
i\hbar\frac{\partial}{\partial t}\left[\begin{array}{c}|\psi_P^{(1)}\rangle \\\vdots \\|\psi_P^{(n)}\rangle\end{array}\right]
=\left[\begin{array}{ccc}\langle\psi_B^{(1)}|\widehat{\mathbf{H}}^{(2)}|\psi_B^{(1)}\rangle & \cdots & \langle\psi_B^{(1)}|\widehat{\mathbf{H}}^{(2)}|\psi_B^{(n)}\rangle \\ \vdots & \ddots & \vdots \\ \langle\psi_B^{(n)}|\widehat{\mathbf{H}}^{(2)}|\psi_B^{(1)}\rangle & \cdots & \langle\psi_B^{(n)}|\widehat{\mathbf{H}}^{(2)}|\psi_B^{(n)}\rangle\end{array}\right]\left[\begin{array}{c}|\psi_P^{(1)}\rangle \\\vdots \\|\psi_P^{(n)}\rangle\end{array}\right],~~i\hbar\frac{\partial}{\partial t} |\psi_B^{(n)}\rangle = \widehat{\mathbf{H}}^{(0)}|\psi_B^{(n)}\rangle,\end{split}
\end{align}
\end{widetext}
which are supplemented with the condition $\langle\psi_B^{(n)}|\psi_B^{(m)}\rangle=\delta_{nm}$. Note that the perturbation vectors $|\psi_P^{(m)}\rangle$'s do not backreact on the background states $|\psi_B^{(m)}\rangle$'s. Nevertheless, they interact with each other through the nondiagonal elements of the above matrix. With this equation we implement the classical condition of the lack of backreaction at quantum level. It reveals the rich physics of primordial fields in quantum cosmological spacetimes.

In what follows we restrict our attention to the simplest product state (\ref{product}) and postpone the study of compound states to future papers.} This is equivalent to the additional assumption that the off-diagonal terms in the matrix (\ref{matrixlaw}) are negligible, and therefore each summand in (\ref{genproduct}) evolves independently according to Eqs (\ref{eom1}) and (\ref{eom11}). The Hamiltonian $\widehat{\mathbf{H}}_p=\langle\psi_B|\widehat{\mathbf{H}}^{(2)}|\psi_B\rangle$ has the following form:
\begin{equation}\begin{split}\label{sham}
\widehat{\mathbf{H}}_p&=\mathcal{S}\sum_{\vec{k}}\frac{1}{2}\left(|\widehat{\pi}_{\pm}(\vec{k})|^2+\Omega^2_k|\widehat{h}_{\pm}(\vec{k})|^2\right),\end{split}
\end{equation}
where $\mathcal{S}=2\mathfrak{g}\langle\left(\frac{\widehat{Q}}{\gamma}\right)^{-2}\rangle$ and $\Omega^2_k=\frac{k^2}{4\mathfrak{g}^2}\frac{\langle\left(\frac{\widehat{Q}}{\gamma}\right)^{\frac{6w+2}{3-3w}}\rangle}{\langle\left(\frac{\widehat{Q}}{\gamma}\right)^{-2}\rangle}$ are determined from the fully quantum background dynamics. The equations of motion for the perturbation variables read
\begin{align}\label{eom}
\frac{1}{\mathcal{S}}\frac{\ud}{\ud t}\left(\frac{1}{\mathcal{S}}\frac{\ud\widehat{h}_{\pm,\vec{k}}}{\ud t}\right)=-\Omega^2_k \widehat{h}_{\pm,\vec{k}}~.
\end{align}
In analogy with the classical case, we introduce a new time parameter and a new dynamical variable, 
\begin{align}\label{eta_t}\begin{split}
\eta=\int\langle\left(\frac{\widehat{Q}}{\gamma}\right)^{-2}\rangle^{\frac{3w-1}{3w-3}}\ud t~,\\
~~\widehat{\mu}_{\pm,k}=\langle\left(\frac{\widehat{Q}}{\gamma}\right)^{-2}\rangle^{\frac{1}{3w-3}}\widehat{h}_{\pm,k},\end{split}
\end{align}
 The above quantum-level definitions lead, as shown below, to the form of the dynamical law that closest resembles the classical counterpart (\ref{wave}). Interestingly, these definitions emphasize the role played by the moment $\langle\widehat{Q}^{-2}\rangle$ that could be viewed as yielding the ``semiclassical" scale factor $ a|_{sem}=\langle\left(\frac{\widehat{Q}}{\gamma}\right)^{-2}\rangle^{\frac{1}{3w-3}} $. It has to be stressed that there are, in principle, infinitely many quantum quantities corresponding to a given classical one and we simply chose the most convenient one. The choice of a new clock and new variables cannot affect the physical predictions of the model. The definitions (\ref{eta_t}) lead to the following { equation of motion}:
\begin{align}\label{wave2}\boxed{
\widehat{\mu}_{\pm,\vec{k}}''+\left(k^2c^2_g-\frac{\left(\langle\widehat{Q}^{-2}\rangle^{\frac{1}{3w-3}}\right)''}{\langle\widehat{Q}^{-2}\rangle^{\frac{1}{3w-3}}}\right)\widehat{\mu}_{\pm,\vec{k}}=0,}
\end{align}
where $c^2_g=\langle\left(\frac{\widehat{Q}}{\gamma}\right)^{\frac{6w+2}{3-3w}}\rangle~\langle\left(\frac{\widehat{Q}}{\gamma}\right)^{-2}\rangle^{\frac{3w+1}{3-3w}}$. Equation (\ref{wave2}) is a quantum version of the gravitational wave propagation equation (\ref{wave}). Comparing it with Eq. (\ref{wave}) we notice that it incorporates two distinct quantum effects on the evolution of gravitational waves. The first effect is due to the quantum term in the background Hamiltonian, which replaces the classical singularity with a bouncing behavior of the expectation values of dynamical variables such as $\langle \widehat{Q}^{-2}\rangle$ or $\langle \widehat{Q}^{\frac{6w+2}{3-3w}}\rangle$. The second effect is due to the quantum uncertainty in the background spacetime. The latter influences both the speed of gravitational waves $c^2_g$ and the interaction potential $V=\frac{\left(\langle\widehat{Q}^{-2}\rangle^{\frac{1}{3w-3}}\right)''}{\langle\widehat{Q}^{-2}\rangle^{\frac{1}{3w-3}}}$. Notice that the uncertainty effect vanishes at the semiclassical level where all the expectation values $\langle \widehat{Q}^{n}\rangle$ are replaced by the respective semiclassical expressions $\langle \widehat{Q}\rangle^{n}$. Moreover, both effects vanish away from the bounce when $\langle \widehat{Q}^{-2}\rangle^{-\frac{1}{2}}$ becomes large, as we show below.

Let us switch to the Heisenberg form for the equation of motion and solve the dynamics for the operator $\widehat Q^2$. Let us first notice the closed algebra of the operators
\begin{equation}
[ \widehat Q^2,\widehat{\mathbf{H}}^{(0)} ] =  4 i \widehat D,~~[\widehat D,\widehat{\mathbf{H}}^{(0)} ]  =  2 i \widehat{\mathbf{H}}^{(0)},~~[\widehat Q^2,\widehat D ]  =  2 i \widehat Q^2,
\label{algebra}
\end{equation}
which allows us to immediately integrate the dynamics
\begin{equation}\label{backSOL}
\widehat D(t) = 2 \widehat{\mathbf{H}}^{(0)} t + \widehat D(0),~~~\widehat Q^2(t) = 4 \widehat{\mathbf{H}}^{(0)} t^2 + 4 \widehat D(0) t + \widehat Q^2(0).
\end{equation}
Thus, for large $|t|$ we find,
\begin{align}\label{infinitepast}\begin{split}
&\lim_{t\rightarrow \pm\infty} c^2_g=\langle\frac{1}{\widehat{\mathbf{H}}^{(0)}}\rangle^{\frac{3w+1}{3-3w}} \langle\widehat{\mathbf{H}}^{(0) \frac{3w+1}{3-3w}}\rangle=const.,\\
&\lim_{t\rightarrow \pm\infty}\frac{\left(\langle\widehat{Q}^{-2}\rangle^{\frac{1}{3w-3}}\right)''}{\langle\widehat{Q}^{-2}\rangle^{\frac{1}{3w-3}}}=0,\end{split}
\end{align}
{ which should be true for any state since $\langle \widehat{\mathbf{H}}^{(0)}\rangle > 0$\footnote{ A heuristic argument can be as follows: The positive self-adjoint operator $\widehat Q^2(t)$ can be viewed as an infinite matrix that can be diagonalized at any moment of time. For large times the elements of this matrix are dominated by the elements of $4 \widehat{\mathbf{H}}^{(0)} t^2$, which has positive eigenvalues too. Therefore, any matrix operator of the form $\widehat Q^{2n}(t)$ should be dominated at large times by the elements of $(4 \widehat{\mathbf{H}}^{(0)} t^2)^{n}$.}.} The above limits show that the oscillation frequency of every mode is asymptotically fixed and well-defined vacuum states for remote past and remote future exists.  

\subsection{Gravitational wave amplitude}
Our convention for the physical dimensions is as follows: the spacetime coordinates are given in units of length, whereas the scale factor, and thus $q$, are dimensionless. The momentum coordinate $p$ has the dimension of mass times length. Analogously, the perturbation variables $h_{\pm}(\vec{k})$ and $\pi_{\pm}(-\vec{k})$ have no dimension and the dimension of mass times length, respectively. 

Let us introduce the annihilation and creation operators,
\begin{align}\label{defaa}\begin{split}
\widehat{h}_{\vec{k}}(t)&=\frac{1}{\sqrt{2}}\left(\widehat{a}_{\vec{k}}h_{k}^*(t)+\widehat{a}_{-\vec{k}}^{\dagger}h_{k}(t)\right),\\
\widehat{\pi}_{\vec{k}}(t)&=\frac{1}{\sqrt{2}}\left(\widehat{a}_{\vec{k}}\frac{1}{\mathcal{S}}\dot{h}_{k}^*(t)+\widehat{a}_{-\vec{k}}^{\dagger}\frac{1}{\mathcal{S}}\dot{h}_{k}(t)\right),\end{split}
\end{align}
where $\widehat{a}_{\vec{k}}$'s are constant, whereas $h_{k}(t)$ are the {\it isotropic} mode functions which solve the {\it isotropic} Eq. (\ref{eom}). Upon setting $h_{k}^*\frac{1}{\mathcal{S}}\dot{h}_{k}-h_{k}\frac{1}{\mathcal{S}}\dot{h}_{k}^*=2i\hbar$, the Hamiltonian $\widehat{\mathbf{H}}_p$ becomes minimal at $t_0$ on the vacuum state $|0\rangle$ such that $\widehat{a}_{\pm,\vec{k}}|0\rangle=0$ if
\begin{equation}\label{ini}
h_{k}(t_0)=\sqrt{\frac{\hbar}{\Omega_{k}(t_0)}},~~\frac{\dot{h}_{k}(t_0)}{\mathcal{S}(t_0)}=i{\sqrt{\hbar\Omega_{k}(t_0)}}.
\end{equation}
{ We actually push the above condition to the infinite past, i.e. $t_0\rightarrow -\infty$, where the spacetime is classical and flat at all cosmological scales as follows from Eq. (\ref{infinitepast}).}

The predictions for primordial gravitational radiation are often given in terms of the amplitude spectrum which is deduced from the equal-time correlation function, where we assume that the interesting coordinate distances $|\vec{x}-\vec{y}|\ll 1$ (or, $k\gg 1$) are small in comparison to the size of the universe. Furthermore, the isotropy $\mu_{\vec{{k}}}=\mu_{{k}}$ is assumed. Following the convention of \cite{Mukhanov:1990me}, we define the spectrum of amplitude of quantum fluctuations of the gravitational waves (per each polarization mode) as
\begin{equation}\label{spdef}
\delta_{\widehat{h}}(k)=\frac{\sqrt{\mathcal{V}_0}}{\langle\left(\frac{\widehat{Q}}{\gamma}\right)^{-2}\rangle^{\frac{1}{3w-3}}}\frac{|\mu_{k}|}{2\pi}k^{\frac{3}{2}}~.
\end{equation}
The amplitude $\delta_{\widehat{h}}(k)$ is time dependent. However, for long-wavelength modes, once their amplitude is set after the bounce, it remains to a large degree constant during a substantial part of the subsequent cosmological evolution. 

{ It is well known that the amplitude (\ref{spdef}) is singular for $k\rightarrow\infty$. The singularity can be removed by means of the adiabatic subtraction \cite{parker}. Nevertheless, we assume that this procedure should not produce any effect at the relevant scales. We provide the asymptotic expansion of the amplitude for the quantum model studied below in Appendix \ref{adiabatic}. }

\section{Internal structure of the bounce}
In this section we argue in favor of the dynamical significance of the inner structure of quantum bounces. We demonstrate the effect of the background wave function on the form of the gravitational wave propagation equation (\ref{wave2}). First, however, we discuss a semiclassical description of the quantum bounce that neglects its inner structure and the interaction potential it leads to.

\subsection{Semiclassical description}\label{semiclassical_description}
\begin{figure}[t]
  \centering
  \includegraphics[width=0.47\textwidth]{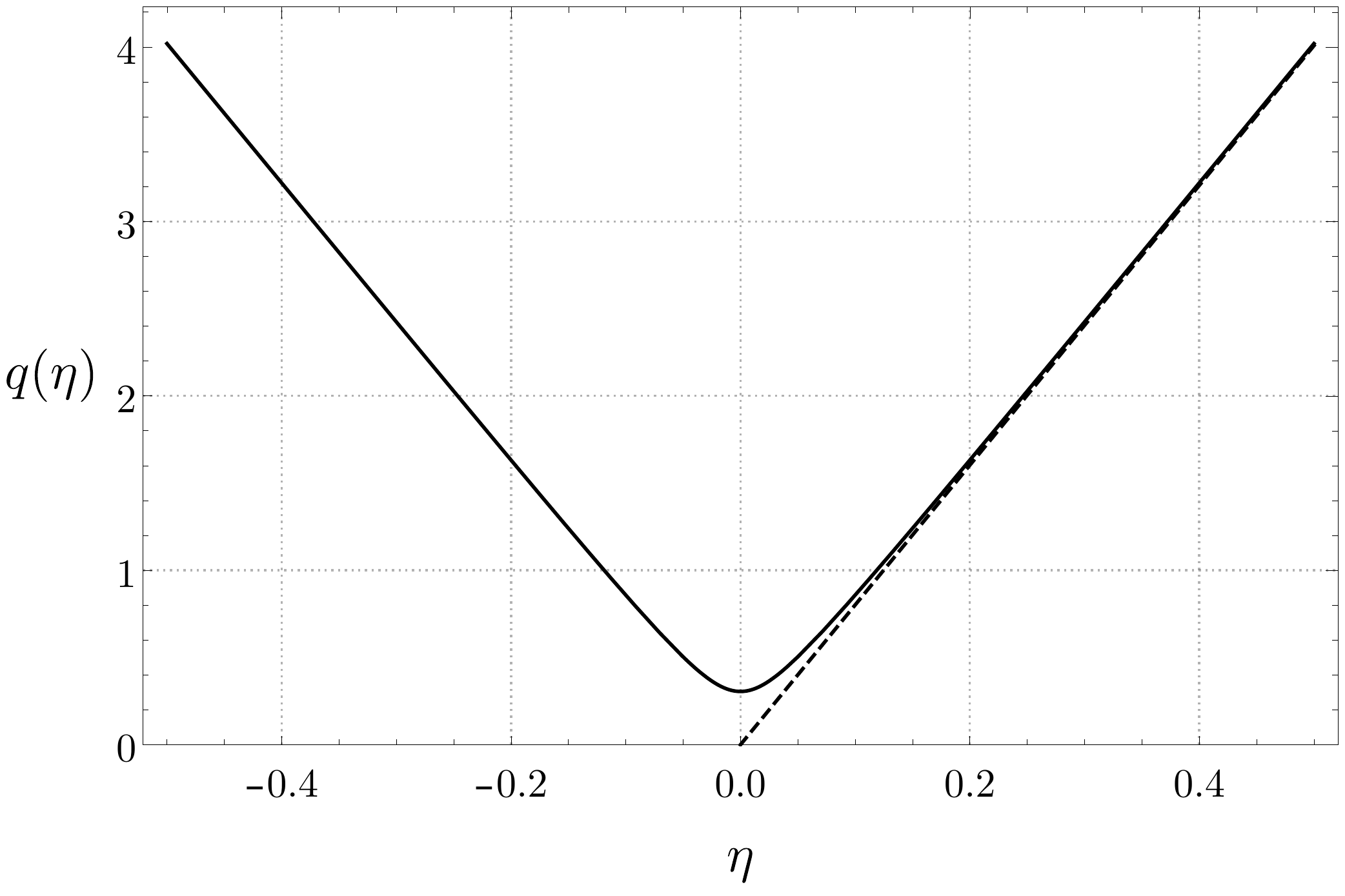}  
\caption{The evolution of $q$ in the classical (dashed line, $K=0$) and semiclassical (solid line, $K=\frac{3}{4}$) background model.}
\label{fig:class_bounce}
\end{figure}

In what follows, we derive the gravitational wave propagation equation (\ref{wave2}) by means of the Ehrenfest equations. Given the dynamics of the expectation values of elementary variables  $\langle \widehat{Q} \rangle(t)$ and $\langle \widehat{P} \rangle(t)$, we form the approximate dynamics of the expectation values of the relevant compound observables. According to the Ehrenfest theorem, the dynamics of the elementary expectation values generated by the zero-order Hamiltonian $\widehat{\mathbf{H}}^{(0)}$ of Eq. (\ref{qham}) reads
\begin{align}\label{eq:ehrenfest}
\frac{d}{dt} \langle \widehat{Q} \rangle = 2 \mathfrak{g} \langle \widehat{P} \rangle, ~~~\frac{d}{dt} \langle \widehat{P} \rangle = 2\mathfrak{g} \hbar^2 K \langle \widehat{Q}^{-3} \rangle.
\end{align}
We assume the probability density in the scale-factor representation,
\begin{equation}\label{eq:ansatz}
\rho(x,t) = \delta(x-q(t)),
\end{equation}
which is a mathematical idealization of a probability density peaked around a semiclassical solution for which $ \langle \widehat{Q} \rangle (t)=q(t)$. It is not an exact solution to the Schr\"{o}dinger equation (\ref{eom1}), but an approximate one, which allows one to immediately  obtain the dynamics of the expectation values of any function of $Q$ once the dynamics of $q$ is known. Discarding the higher moments of $Q$ in the wave function may only be temporarily a valid approximation due to the natural spreading of the probability distribution with time.\\
We find the solution to (\ref{eq:ehrenfest}) to read
\begin{subequations}
\begin{align}\label{subeq:x}
\langle \widehat{Q} \rangle(t) &= q_b\sqrt{(k_{max}t)^2+1}, ~\\
\langle \widehat{P} \rangle (t) &= \frac{q_b k_{max}^2t}{2\mathfrak{g}\sqrt{(k_{max}t)^2+1}},
\end{align}
\end{subequations}
where $q_b=\sqrt{\frac{2\mathfrak{g}\hbar^2K}{\mathbf{H}^{(0)}_{sem}}}$, $k_{max}=\frac{\mathbf{H}^{(0)}_{sem}}{\hbar\sqrt{K}}$, and $\mathbf{H}^{(0)}_{sem}= \mathfrak{g} \left(\langle \widehat{P} \rangle^2 + \frac{\hbar^2 K}{\langle \widehat{Q} \rangle^2}\right)$ is assumed to be equal to the classical value $\mathbf{H}^{(0)}_{sem}=\mathbf{H}^{(0)}$. We plot a typical solution in Fig. \ref{fig:class_bounce}. The classical and semiclassical trajectories are the same away from the singularity, which proves the correct behavior of the semiclassical model. Close to the singularity, the classical and semiclassical trajectories diverge as the former terminates (or originates) in the singularity, whereas the latter avoids the singularity through a bounce.\\

{ We assume the quantum universe to be filled with a fluid with an effective equation of state, which at some point of cosmological expansion becomes radiation so that a connection with the observable universe can be made.} Making use of the relations below Eq. (\ref{defs}) we obtain the scale factor of the universe at the bounce, from which we infer the redshift at the bounce,
\begin{align}
z_b= \left(\frac{10^{120}r}{(1-w)\sqrt{K}}\right)^{\frac{2}{3(1-w)}}{z_T}^{\frac{\frac{1}{3}-w}{1-w}},
\end{align}
where $z_T$ is the { fluid-to-radiation transition redshift}. If the early Universe had not undergone the fluid transition, then 
we have $z_b\approx 10^{120}\frac{r}{\sqrt{K}}$ and the physical wavelength of the pivot mode at the bounce  reads $\lambda_{*,phys}\approx 5\cdot 10^{-60}\frac{\sqrt{K}}{r}l_P$. This implies a huge value of $\frac{\sqrt{K}}{r}$ for a cosmological scenario in which the observable cosmological scales are around the order of $l_P$ at the bounce. In general, we note that the bigger the universe is and the more energy it contains, the smaller the volume at which it bounces. The inverse is true for the value of $\sqrt{K}$. Because the amount of energy in the observable universe is so huge, the quantum correction preventing the singularity comes to dominate the dynamics at the Planck volume only if the value of $\sqrt{K}$ is very large. Nevertheless, we may fine-tune the model to yield a bounce exactly at Planck scale.\\

We shall now turn to the evolution of the coefficients $c_g^2(t)$ and $V(t)$ in the gravitational wave propagation equation (\ref{wave2}).
It is straightforward to obtain them in the current approximation,
\begin{subequations}
\begin{align}
&c_g^2(t)\biggr\rvert_{sem} = \frac{q^{\frac{6w+2}{3-3w}}(t)}{q^{\frac{6w+2}{3-3w}}(t)} = 1, \label{subeq:cgclass}\\\begin{split}
&V(t)\biggr\rvert_{sem} = \frac{q''(t)}{q(t)}=k_{max}^2\bigg[\frac{q_b^2}{\gamma^2}(1+(k_{max}t)^2)\bigg]^{\frac{6w-2}{3w-3}}\\
&\times\frac{(2-6w)(k_{max}t)^2+(6-6w)}{(3w-3)^2[1+(k_{max}t)^2]^2},\label{subeq:Vclass}\end{split}
\end{align}
\end{subequations}
where prime $'$ denotes differentiation with respect to conformal time $\eta$ defined in Eq. (\ref{eta_t}). We note that the typical length scale influenced by the bounce is given by the factor $\big(\frac{q_b}{\gamma}\big)^{\frac{6w-2}{3w-3}}k_{max}$ and thus, we introduce a dimensionless quantity $\tilde{k}=\big(\frac{q_b}{\gamma}\big)^{\frac{6w-2}{3-3w}}\frac{k}{k_{max}}$ to express the scale dependence of the gravity-wave amplitude. Similarly, the typical timescale at which the bounce operates is given by $k_{max}$; hence we introduce a dimensionless quantity $\tilde{t}=k_{max}t$.

\begin{figure}[t]
    \centering
    \includegraphics[width=0.47\textwidth]{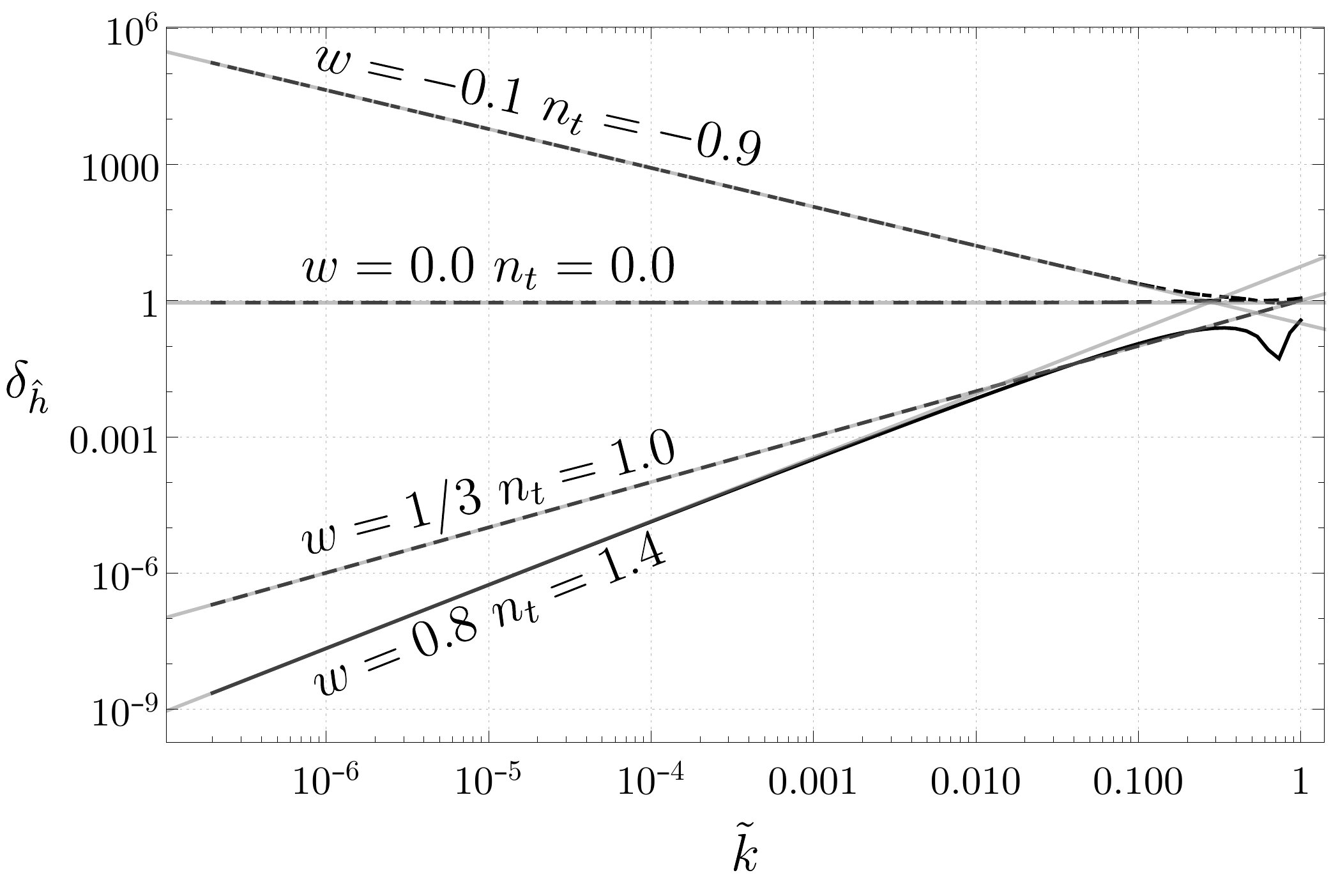}
       \caption{The primordial amplitude spectrum $\delta_h(\tilde{k})$ in a semiclassical universe with a big bounce and a cosmological fluid for a few values of $w=\frac{\rho}{p}$.}
       \label{n_t}
\end{figure}

\begin{figure}[t]
    \centering
    \includegraphics[width=0.47\textwidth]{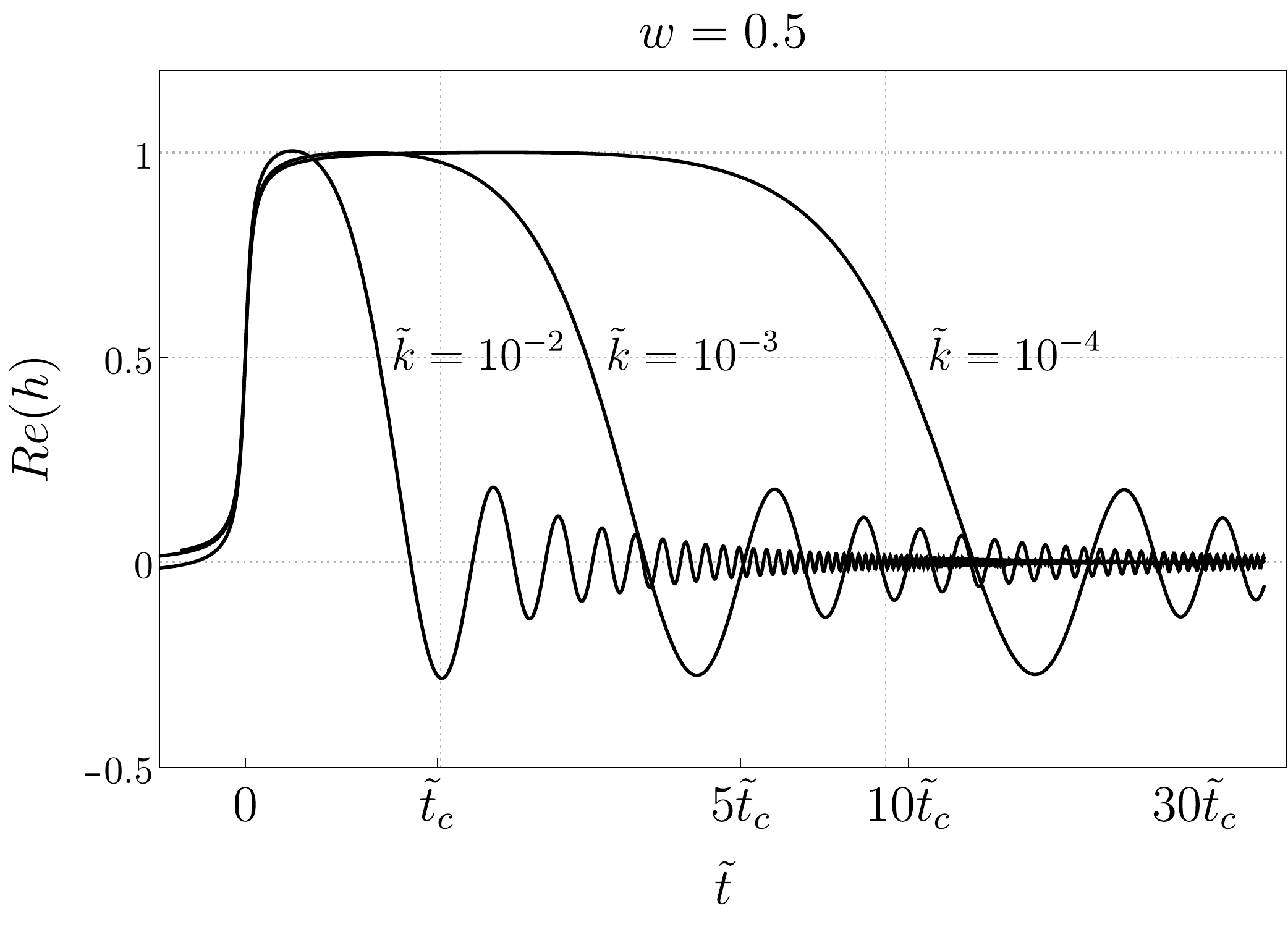}
       \caption{The evolution of the amplitude of a few modes in a semiclassical universe with a big bounce and a cosmological fluid with $w=0.5$. For clarity the maximum amplitude of h has been normalized to unity.}
       \label{mode}
\end{figure}

In the semiclassical treatment of Eq. (\ref{wave2}), the gravitational waves propagate at the speed of light and { interact with the potential (\ref{subeq:Vclass}) induced by the evolution of the universe}. A similar potential was obtained within the Bohm-de Broglie approach in \cite{Peter2006} where the long-wavelength amplitude spectrum was found to have the spectral index $n_{t}=\frac{6w}{1+3w}$.\footnote{Authors of \cite{Peter2006} consider the spectral index of the power spectrum rather than the amplitude spectrum, hence the difference by factor $2$.} In Fig. \ref{n_t} we provide an independent verification of their result by numerical integration of the primordial amplitude for a few values of $w$ and a range of modes $\tilde{k}$. The analytical computation of the primordial spectrum for this and other interaction potentials is discussed in Sec. \ref{lwa}. The time evolution of a few modes of the primordial gravitational wave is plotted in Fig. \ref{mode}. 

Let us describe the relation between the primordial amplitude and the value of $K$ and $w$. One might think that since the larger the value of $K$ the less redshifted and milder the bounce is, the amplitude should decrease as $K$ increases. However, it can be shown that the amplitude scales with $K$ as $A_t~\propto~ K^{\frac{5w-1}{2(1-w)}}$ (see Sec. \ref{lwa} for the explicit formulas). It follows that, for $w<\frac{1}{5}$, the larger the value of $K$ (and the stronger the quantum effect) the smaller the primordial amplitude as one would expect. On the other hand, when $w>\frac{1}{5}$, this relation becomes inverted; that is, the larger the value of $K$, the larger the primordial amplitude. Hence, respecting the upper bound on the amplitude one may decrease the value of $K$ as much as one wishes for $w>\frac{1}{5}$. Note that the case $w=\frac{1}{5}$ is a borderline for which the primordial amplitude does not actually depend on $K$. Its value for $w=\frac{1}{5}$, $z_T=10^{28}$, and $r=2$ reads $A_t\approx 10^{14.6}.\footnote{The value $z_T=10^{28}$ corresponds to the ``end-of-inflation" redshift \cite{dodelson}.}$

\begin{figure}[t]
    \centering
    \includegraphics[width=0.47\textwidth]{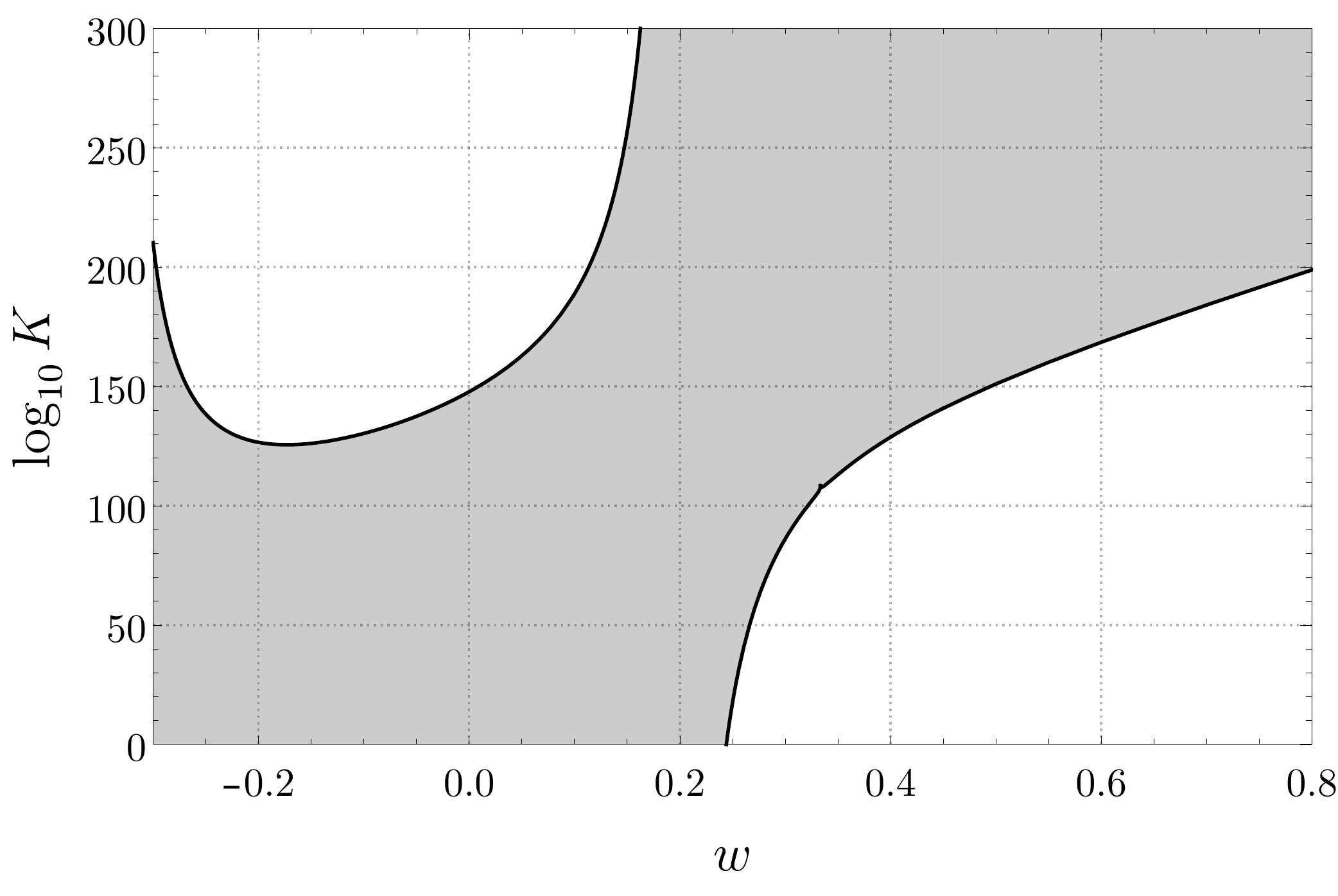}
       \caption{The white regions represent the admissible values of the parameter $K$ as functions of $w$ ($r=2$, $z_T=10^{28}$).}
       \label{Kw}
\end{figure}

\begin{figure}[t]
    \centering
    \includegraphics[width=0.47\textwidth]{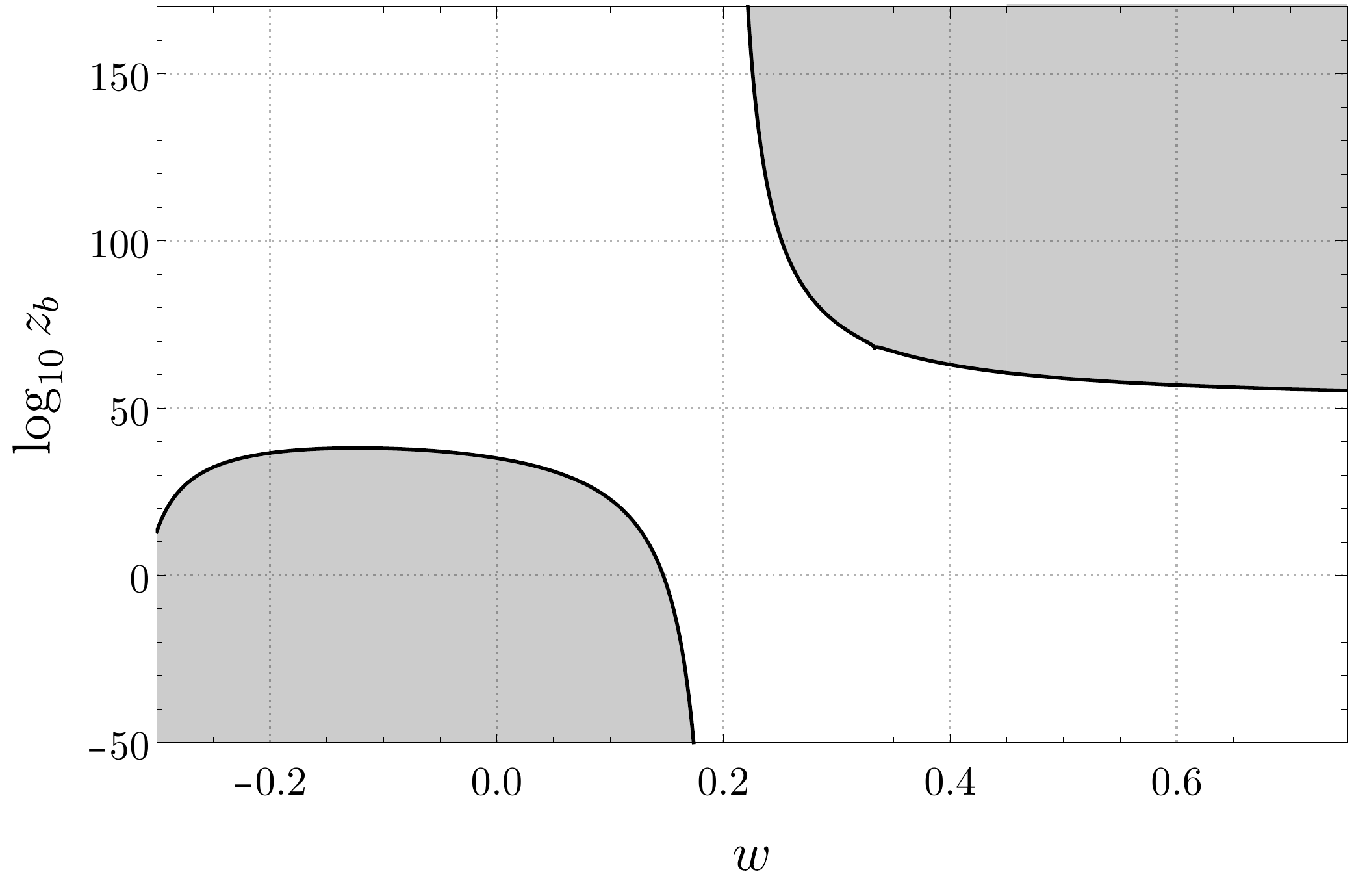}
       \caption{The white region represents the admissible values of the bounce redshift $z_b$ as functions of $w$ ($r=2$, $z_T=10^{28}$).}
       \label{zw}
\end{figure}

\begin{figure}[t]
    \centering
    \includegraphics[width=0.47\textwidth]{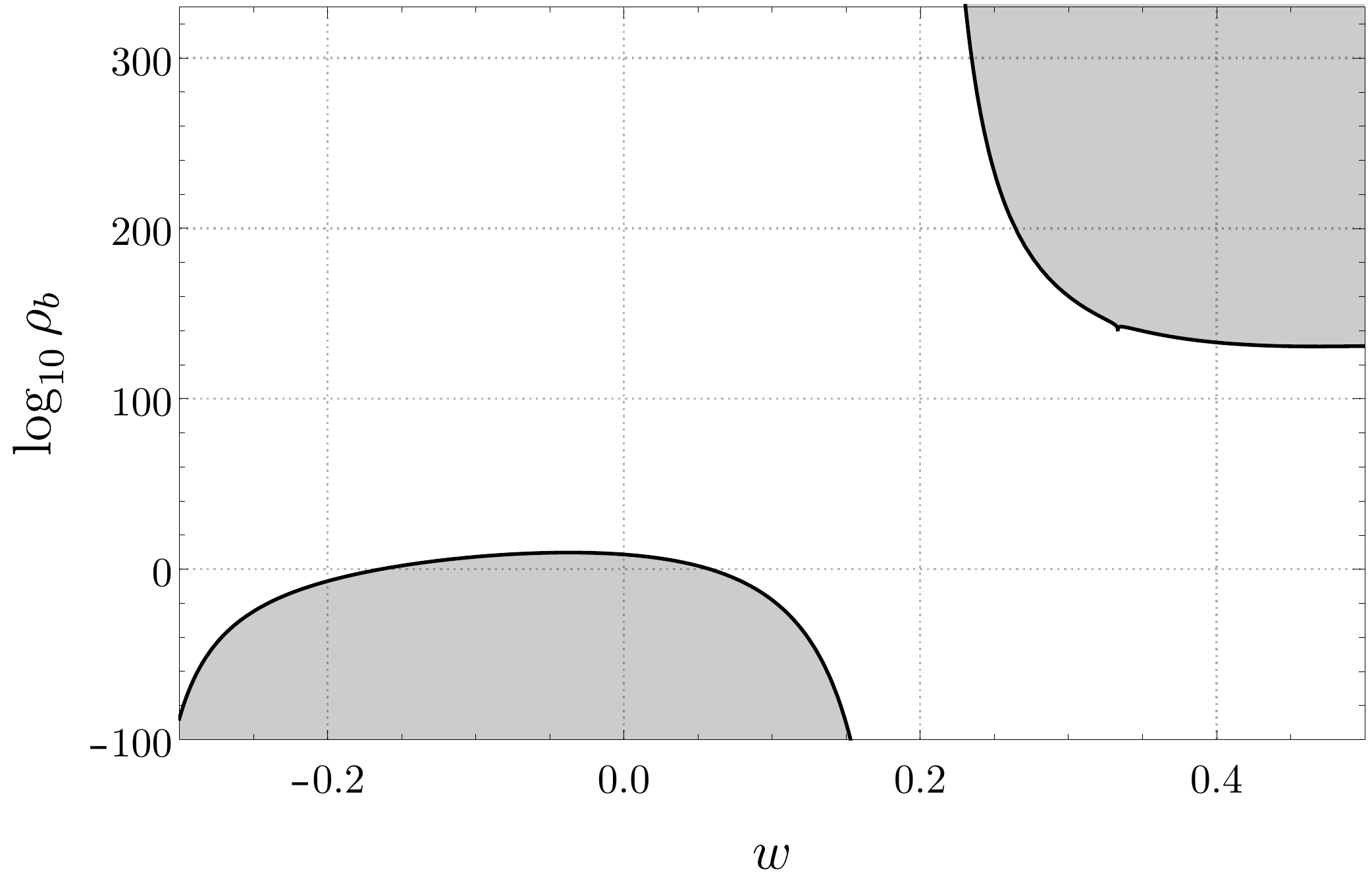}
       \caption{The white region represents the admissible values of the bounce energy density $\rho_b$ as functions of $w$ ($r=2$, $z_T=10^{28}$).}
       \label{rhow}
\end{figure}

Let us assume that the gravitational wave amplitude at the pivot scale should not exceed $10^{-5}$, to be consistent with the Planck data for $k_*=0.002$Mpc$^{-1}$ \cite{cmb}. This in turn puts constraints on the free parameter $K$. In Fig. \ref{Kw} we plot the required value of $K$ as a function of the fluid's type. We find huge values allowed for almost all $w$'s. In Fig. \ref{zw} we plot the redshift (and the energy density in Fig. \ref{rhow}) at the bounce if the pivot scale amplitude reads $10^{-5}$. These results clearly suggest that for the modes of interests we may avoid the so-called trans-Planckian problem as the observable modes when propagating through the bounce, where they are the shortest, may exceed the Planck length by a number of orders of magnitude. On the other hand, they indicate that there might be the problem of unnaturally large value of $K$. On the grounds that it is a quantum correction one expects that it should be of order of unity in Planck units while it has to be of many, many orders of magnitude larger for $w<\frac{1}{5}$ in order to produce $A_t=10^{-5}$.

There are two ways to argue for the possibility of a large $K$ in our quantum model. Both arguments refer to the ways in which we think about quantization of gravitational systems. First, note that we do not know which choice of basic variables is correct for quantization of gravitational systems. In the preceding section we chose dilation $\hat{D}$ and position $\hat{Q}$ but we did not specify the ordering of these operators in the Hamiltonian. In \cite{kasner} it was actually shown that $-\frac{1}{4}<K<\infty$ depending on the chosen ordering. Thus, large values of $K$ can be easily accommodated by theory. The second argument is more subtle and is based on the nature of dynamics in quantum gravity. It is known (see e.g. \cite{FRW} and references therein) that quantities like the scale of the bounce are not physically meaningful (or, unambiguous) in quantum gravity unless one indicates the internal clock used for computing those quantities. This property is referred to as the time problem. It follows that the scale of the bounce obtained in the present model is tied to the specific choice of clock $t$ that we have made for the derivation of the model. One might have chosen another clock and found much more Planckian, or even sub-Planckian, scale of the bounce issued from a weaker repulsive potential, i.e. a smaller value of $K$. The contradiction between those two conclusions would be, however, only illusory as it was shown in \cite{BI} that the physical predictions for the classical phase of the cosmological evolution derived from both models must agree with each other, as for instance, in regard to the predicted value of the amplitude of primordial gravitational waves in a large expanding universe. Finally, let us note that the value of $K$ allowed by the cosmological observations can be extremely large \cite{szydlowski}.

Above we have derived and described the semiclassical model. It remains to verify whether adding a substantial amount of quantum spread to the cosmological background can alter the model in some important ways, in particular, whether the final gravity-wave amplitude is modified in this case due to some modifications of the gravity-wave propagation speed (\ref{subeq:cgclass}) or modifications of the interaction potential (\ref{subeq:Vclass}). We shall investigate this issue below.\\

\subsection{Quantum description}\label{quantum_description}
The mathematical idealization of the probability density made above yields immediately the quantum dynamics of the universe with the classical behavior for large volumes. The interaction potential issued from such an approximation seems rather universal as it was also found in another trajectory approach \cite{Peter2006}. Trajectories are the usual way in which quantum cosmological bounces are described. However, this description completely neglects the quantum uncertainty in the numerical values for the size and the expansion rate of the universe close to the bounce. It is legitimate to ask whether the amount of uncertainty that is completely negligible for the presently large universe might have played a significant role when the universe was small. The nonvanishing uncertainty should be reflected in the dynamics of the coefficients $c_g^2 (t)$ and $V(t)$ of Eq. (\ref{wave2}) as they depend on higher-order moments. As a result, the primordial structure and gravitational waves could be influenced by this purely quantum effect.

\begin{figure}[t]
    \centering
    \includegraphics[width=0.44\textwidth,height=0.4\textwidth]{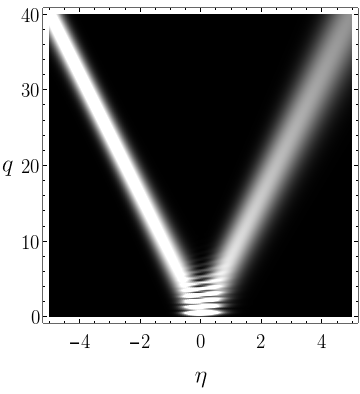}
       \caption{The evolution of the density distribution in $q$ yielded by the exact wave packet (\ref{full_solution}) with $p_0=-4$, $x_0=30$, and $\sigma=2$ ($K=\frac{3}{4}$).}
       \label{wavefunction}
\end{figure}

In what follows we solve the complete dynamics of the background model without any approximation and plot the resulting interaction potential. The analytically integrable solutions are very few and they require numerical integration of the expectation value of $\hat{Q}^{-2}$. We use an analytical three-parameter solution to the background Schr\"odinger equation (\ref{eom1}),
\begin{align}\label{full_solution}\begin{split}
&\langle q | \psi_B \rangle ~ \propto ~ \frac{\sqrt{q}\sigma e^{ - \frac{1}{\sigma^2 + it} \left( ip_0^2 \sigma^2 t + p_0 q_0 t - q^2/4- q_0^2/4 \right) }}{\sqrt{2\pi} (\sigma^2+i t) } \\ 
&\times ~ I_{\sqrt{K+\frac{1}{4}}} \left( \frac{q(2 i  p_0 \sigma^2 +  q_0)}{ 2(\sigma^2+it) }  \right),\end{split}
\end{align}
where { $I_{n}(x)$ is the modified Bessel function of the first kind,} $\hbar=1=\mathfrak{g}$, and $q_0$, $p_0$, and $\sigma$ are free parameters. The evolution of the associated density distribution is plotted in Fig. \ref{wavefunction}. We see a wave packet moving toward the boundary $q=0$ and strongly self-interfering as it bounces against the repulsive potential. The spread of the wave packet is growing as it moves away from the boundary.

\begin{figure}[t]
    \centering
    \includegraphics[width=0.47\textwidth]{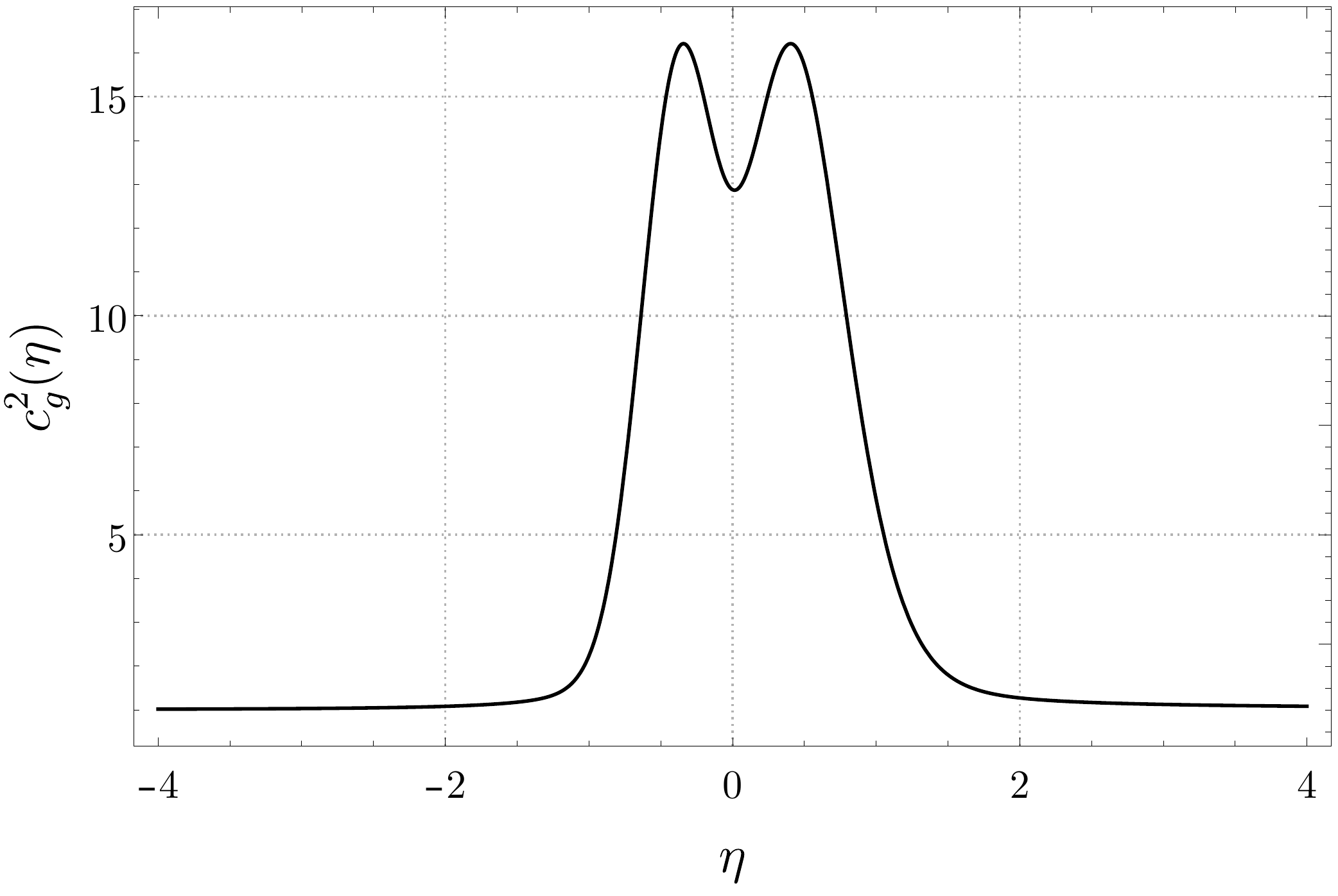}
       \caption{The speed of gravitational waves squared $c^2_g$ of the propagation equation (\ref{wave2}) generated by the background dynamics in the analytical state (\ref{full_solution}). We set $p_0=-4$, $q_0=30$, $\sigma=2$, $K=\frac{3}{4}$.}
                 \label{cg}
\end{figure}

\begin{figure}[t]
    \centering
    \includegraphics[width=0.47\textwidth]{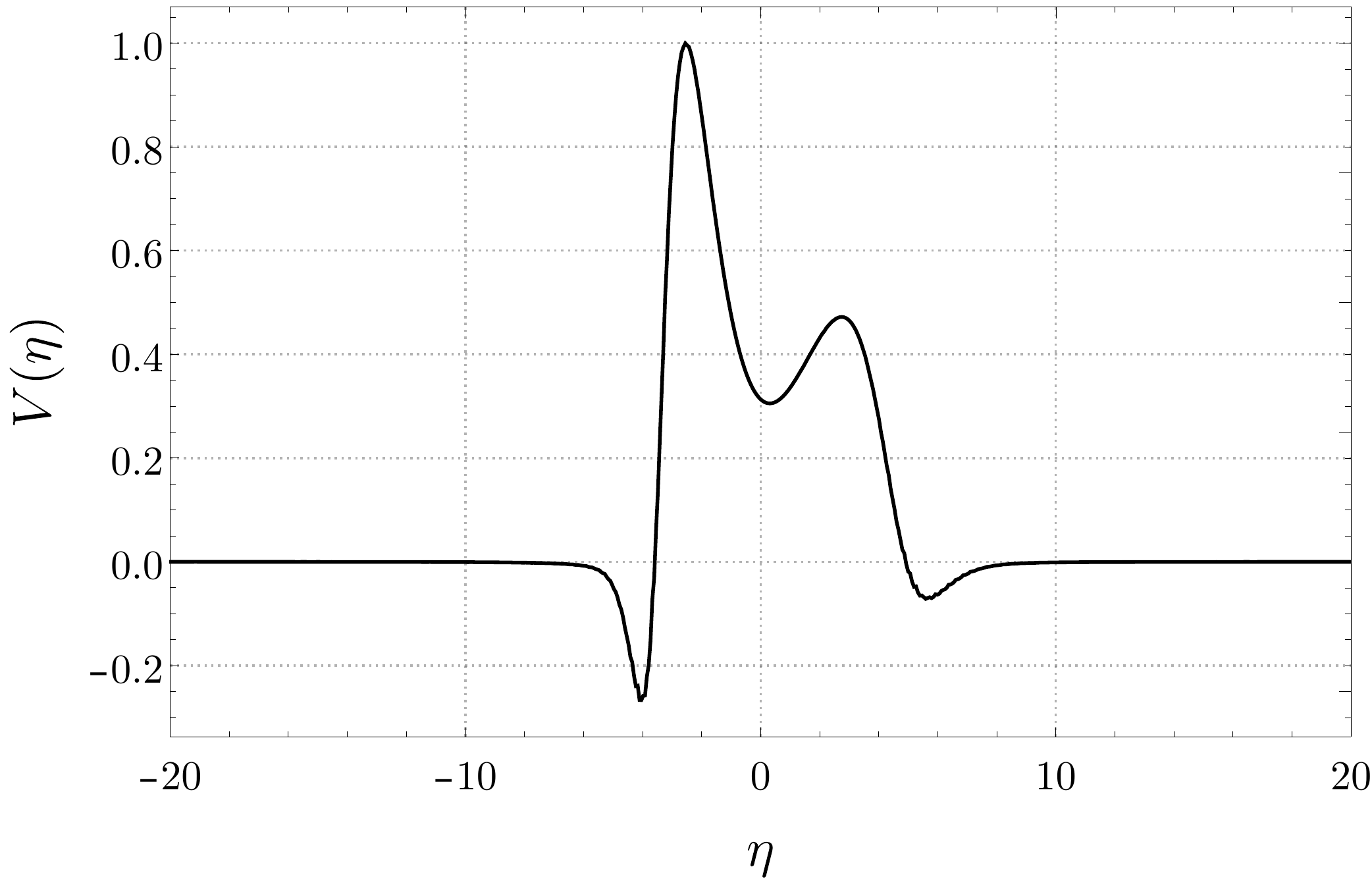}
       \caption{The interaction potential $V$ of Eq. (\ref{wave2}) issued from the fully quantum background dynamics described by (\ref{full_solution}). We set $p_0=-4$, $q_0=30$, $\sigma=2$, $K=\frac{3}{4}$,$w=\frac{1}{3}$.}
                 \label{fullV}
\end{figure}

The evolutions of the coefficients $c_g^2(t)$ and $V(t)$ obtained from the solution (\ref{full_solution}) are plotted in Figs \ref{cg} and \ref{fullV}, respectively. The speed of waves squared $c_g^2(t)$ consists of two maxima separated by a minimum exactly at the bounce and it rapidly decreases to the value $c_g^2=1$ as $t \rightarrow \pm \infty$. The fact that $c_g^2(t)\geqslant 1$ follows from the Schwarz inequality. The brief decline in $c_g^2$ exactly at the moment of the bounce is due to the momentary reduction of the spread as the wave packet bounces off the potential. The fact that $c_g^2$ becomes larger than unity is interpreted as the breakdown of  the "semiclassical spacetime" interpretation of the model rather than as a superluminal propagation of the gravitational waves. We do not expect, however, a significant influence of the dynamical $c_g^2$ on the amplification of long-wavelength gravitational waves precisely because they are assumed to satisfy $k^2\ll V$ at the bounce and the term $\propto~ k^2$ in Eq. (\ref{wave2}) is simply negligible.\\

Similarly, the interaction potential $V(t)$ in Fig. \ref{fullV} displays an extra structure that does not occur for the semiclassical solution given by Eq. (\ref{eq:ansatz}).\\

\subsection{WKB approximation}\label{lwa}

The available analytical solutions do not allow for obtaining an analytical formula for the interaction potential $V$ and the numerical integration of $V$ is cumbersome for large $K$. Therefore we resort to the WKB approximation \cite{Dirac}. We assume the solution to the Schr\"odinger equation (\ref{eom1}),
\begin{align}\label{wkb}
\langle x|\psi_B\rangle(t)=A(x,t)\exp{[iS(x,t)/\hbar]}, ~~A,S\in\mathbb{R},
\end{align}
which when expanded in $\hbar$ yields at lowest order,
\begin{align}
\partial_{t}S=-\mathfrak{g} \left(S_{,x}^2 + \frac{\hbar^2 K}{x^2}\right),~~~\partial_{t}A^2=-2\mathfrak{g}\partial_x(A^2S_{,x}),
\end{align}
where $S$ is the Hamilton's principal function
\begin{align}
S(t,x)=\mathfrak{g}\int^{t,x}\left(\frac{1}{4}x^2_{,t'}-\frac{K}{x^2}\right)\ud t',
\end{align}
where the integral is taken over the semiclassical trajectories with fixed initial condition and $A^2$ behaves like the density of particles following the semiclassical trajectories.\\
Let us assume the probability distribution at the moment of the bounce to read
\begin{align}
\rho(x,0)=\rho(x).
\end{align}
The solution (\ref{wkb}) reads now
\begin{align}
\langle x|\psi_B\rangle(t)=\sqrt{\rho(x_b(x,t))\frac{\partial x_b}{\partial x}}\cdot\exp{[iS(x,t)/\hbar]},
\end{align}
where
\begin{align}\begin{split}
x_b^2(x,t)&=\frac{1}{2}\left(x^2+\sqrt{x^4-16\mathfrak{g}^2\hbar^2Kt^2}\right),\\
S(x,t)&=\frac{\mathfrak{g}^3\hbar^2K}{x_b^2(x,t)}t-\frac{\sqrt{K}(1+\mathfrak{g}^2\hbar^2)}{2\hbar}\arctan{\left(\frac{2\mathfrak{g}\hbar\sqrt{K}}{x_b^2(x,t)}\right)}.\end{split}
\end{align}
Within the WKB approximation the sought expectation value $\langle \widehat{Q}^{-2}\rangle$ reads
\begin{align}\label{rhoWKB}\begin{split}
\langle \widehat{Q}^{-2}\rangle(t)&=\int_0^{\infty}x^{-2}\rho(x_b(x,t))\frac{\partial x_b}{\partial x}\ud x\\
&=\int_0^{\infty}\frac{\rho(x_b)\ud x_b}{x_b^2+\frac{4\mathfrak{g}^2\hbar^2K}{x_b^2}t^2},\end{split}
\end{align}
where in the last line we switched to the Heisenberg picture. The formula (\ref{rhoWKB}) yields an analytical expression for some choices of $\rho(\cdot)$ and thereby it yields an analytical expression for the interaction potential $V$. 

\begin{figure}[t]
    \centering
    \includegraphics[width=0.47\textwidth]{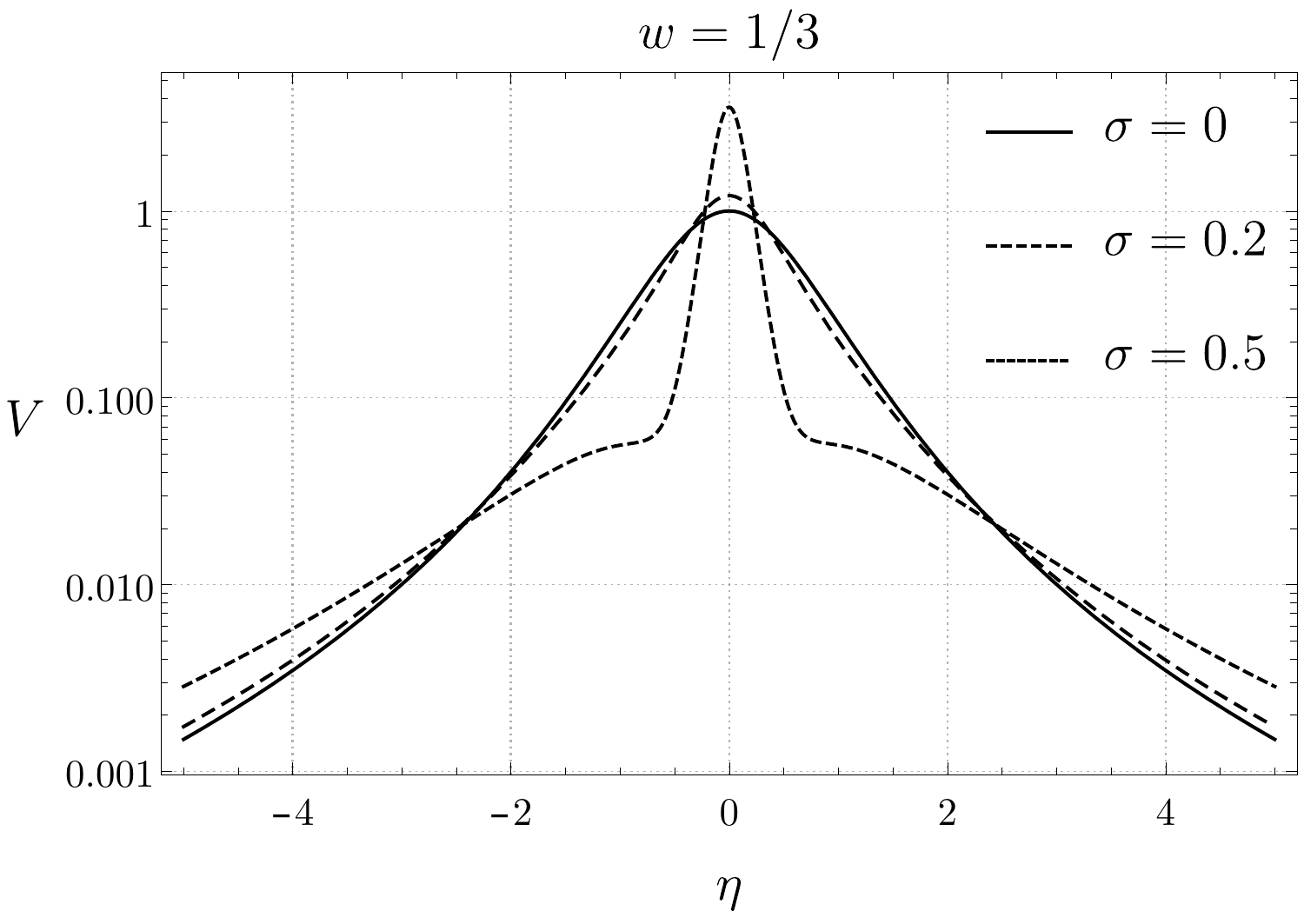}
       \caption{The interaction potential $V$ in the semiclassical (dispersion $\sigma=0$) and the WKB approximation (dispersion $\sigma=0.2$ and $\sigma=0.5$).}
       \label{wkbv}
\end{figure}

Let us assume the density distribution at the bounce to read
\begin{align}\label{density}
\rho(x)=\frac{x}{2q_b^2\sigma}\chi_{[q_b(1-\sigma),q_b(1+\sigma)]}(x),
\end{align}
where $\chi_{[q_b(1-\sigma),q_b(1+\sigma)]}(x)$ is the characteristic function, $q_b$ is a fixed bouncing point and $0<\sigma<1$ is a free dimensionless parameter. We then find
\begin{align}\begin{split}\label{wkbext}
\langle \widehat{Q}^{2}\rangle(t)&=q_b^2\left(1+\sigma^2+\frac{\ln\big|\frac{1+\sigma}{1-\sigma}\big|}{2\sigma}(k_{max}t)^2\right),\\
\langle \widehat{Q}^{-2}\rangle(t)&=\frac{1}{8q_b^2\sigma}\ln\bigg|\frac{(1+\sigma)^4+(k_{max}t)^2}{(1-\sigma)^4+(k_{max}t)^2}\bigg|,\\
\langle \widehat{Q}\rangle\big|_{t=0}&=q_b,\\
(\Delta \widehat{Q})^2\big|_{t=0}&=\langle \widehat{Q}^{2}\rangle-q_b^2=q_b^2\sigma^2.
\end{split}
\end{align}
It follows from the last equality that $\sigma$ has the interpretation of the relative volume dispersion. Notice that for $\sigma\rightarrow 0$ one naturally retrieves the semiclassical description of Sec. \ref{semiclassical_description} as
\begin{align}\begin{split}
\rho(x)&\rightarrow\delta(x-q_b),\\
\langle \widehat{Q}^{2}\rangle(t)&\rightarrow q_b^2\left(1+(k_{max}t)^2\right),\\
\langle \widehat{Q}^{-2}\rangle(t)&\rightarrow\frac{1}{q_b^2(1+(k_{max}t)^2)},\\
(\Delta \widehat{Q})^2\big|_{t=0}&\rightarrow 0.\end{split}
\end{align}
Hence, Eqs (\ref{wkbext}) provide a one-parameter extension to the semiclassical model with the free parameter being the spread of the wave function. The resultant formulas for the interaction potential and the speed of gravitational waves are given in Appendix \ref{wkbapp}.\\

The interaction potential $V$ in time $\tilde{t}=k_{max} t$ produced by the density distribution (\ref{density}) is plotted  for a few values of $\sigma$ in Fig. \ref{wkbv}. It is apparent that the potential is very sensitive to the value of  $\sigma$. In fact, the height and the width of its peak can be altered by many orders of magnitude by the spread of the wave packet. Thus, it is natural to expect that the amplitude spectrum generated by these potentials can also be substantially altered.

\begin{figure}[t]
    \centering
    \includegraphics[width=0.47\textwidth]{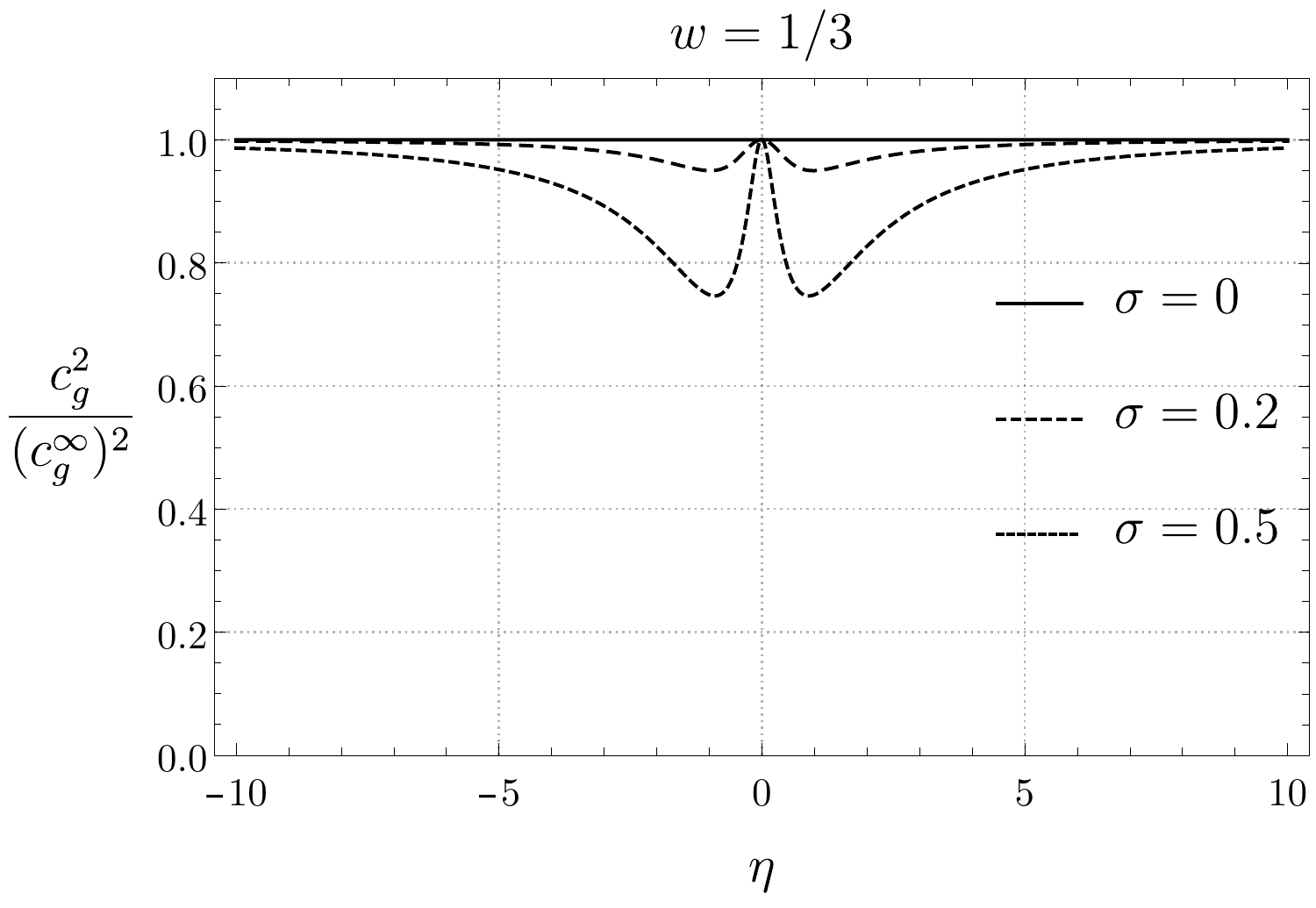}
       \caption{The gravity-wave speed in the semiclassical approximation (dispersion $\sigma=0$) and in the WKB approximation (dispersion $\sigma=0.2$ and $\sigma=0.5$).}
       \label{wkbc}
\end{figure}

In Fig. \ref{wkbc} we plot the evolution of the { renormalized} speed of gravitational waves. For the studied WKB states ($\sigma>0$), the speed decreases just before the bounce and then at the bounce it increases back to its asymptotic value. It behaves symmetrically in time after the bounce. As before, we interpret this behavior as a breakdown of the semiclassical interpretation {of dynamics}. This behavior suppresses the value of the $k^2$ term in Eq. (\ref{wave}) precisely at the moment when it is already subdominant and therefore, the dynamical effect of the varying speed of gravitational waves is negligible. The speed has been normalized so that it asymptotically converges to unity. The fact that it { asymptotically} converges to a different value than unity is not physically relevant as this discrepancy is removed by simply redefining the length scale so that the measured wave number is $k_{eff}=c^{\infty}_gk$, where $c^{\infty}_g$ is the asymptotic value of $c_g$. In this way, the measured speed of gravitational waves equal to unity is retrieved.\\

\begin{figure}[t]
    \centering
    \includegraphics[width=0.47\textwidth]{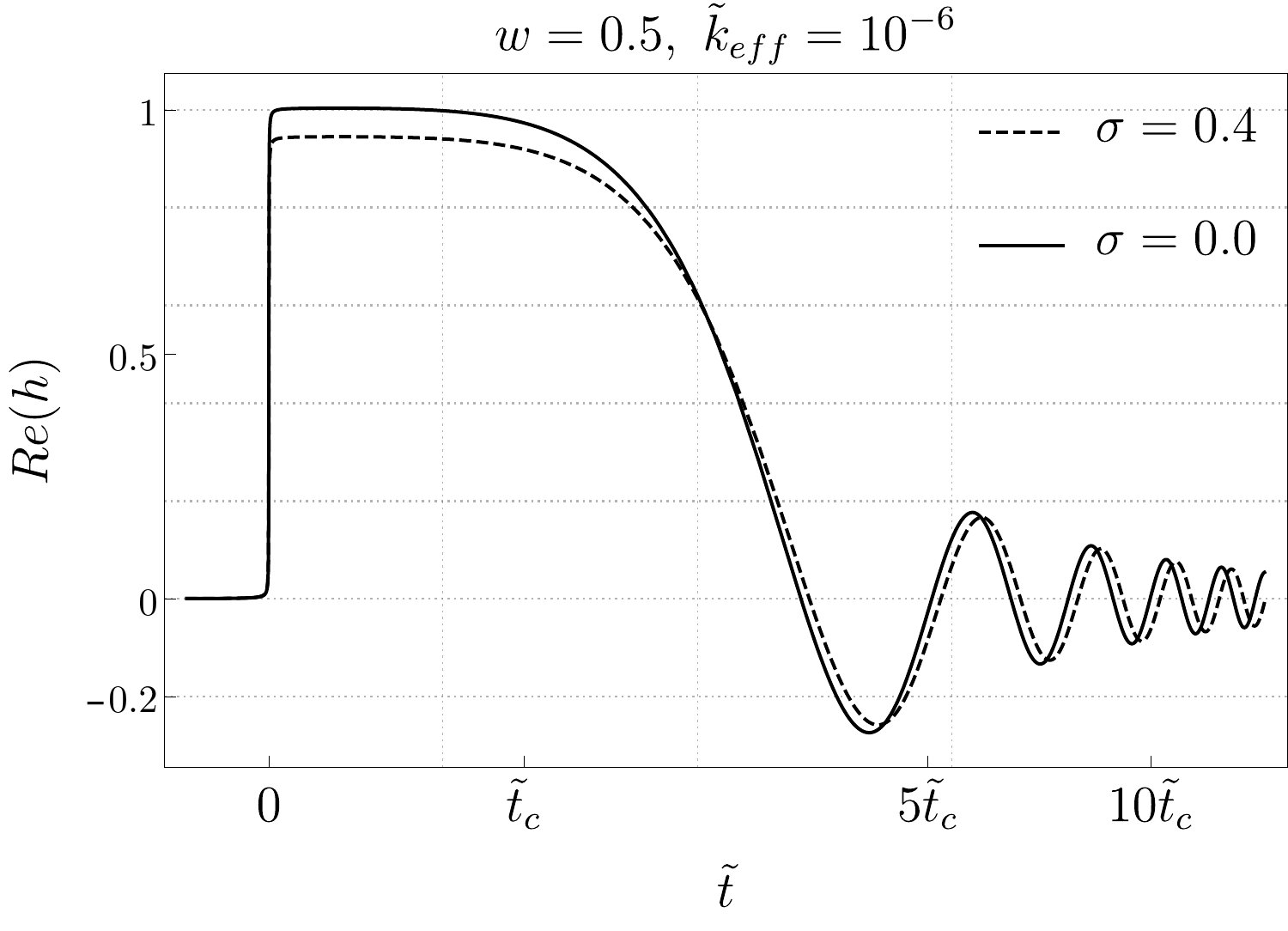}
       \caption{The evolution of a selected mode in the semiclassical (dispersion $\sigma=0$) and the WKB (dispersion $\sigma=0.4$) approximation.}
       \label{mode2}
\end{figure}

In Fig. \ref{mode2} we plot the numerically integrated evolution of the amplitude of a selected mode both in the semiclassical and the WKB approximation. We obtain a noticeable suppression of the amplitude in the WKB approximation.

Now we turn to the analytical computation of the primordial amplitude spectrum for the interaction potentials issued from the WKB approximation. We solve the wave propagation equation (\ref{wave2}) by employing the piecewise approximation used also in \cite{Peter2006}.\footnote{See, in particular, Sec. IV B in \cite{Peter2006}: ``Piecewise approximation and matching in the flat spatial section case."} Below we restrict our discussion to a sketch of the derivation, details of which are found in Appendix \ref{Appendix:spectrum}.\\
 
In agreement with our result for the semiclassical case, we assume that the modes of interest are much longer than the length scale imposed by the interaction potential, $\tilde{k}^{-1} \gg 1$. Such modes cross the potential when $c_g^2k^2=V$ at times $-\tilde{t}_c$ and $\tilde{t}_c$ much larger than the timescale imposed by the interaction potential, $\tilde{t}_c \gg 1$. Therefore, we assume that for $|\tilde{t}| \geqslant\tilde{t}_c$ the potential $V$ is completely classical. This will simplify the evaluation of the solution in this evolution regime. Moreover, since the speed of gravitational waves $c_g$ in the WKB approximation does not converge to unity for large $|\tilde{t}|$'s, we use the effective wave number, $\tilde{k}_{eff} = c_g^{\infty} \tilde{k}$ and $k_{eff} = c_g^{\infty} k$, where $c_g^{\infty}=\lim_{|\tilde{t}| \to \infty} c_g$. It is to be stressed that only the effective quantities are measurable.\\

The dynamics is solved separately in two distinct evolution regimes. The first already mentioned regime spans from the remote past up to the moment when a particular mode crosses the interaction potential, $-\tilde{t}_c$. In this regime we solve the wave equation (\ref{wave2}) in its asymptotic form,
\begin{equation}\label{asymptotic_wave2}
\widehat{\mu}_{\pm,k}''+\left(\left(c_g^{\infty}k\right)^2+ \frac{2(3w-1)}{(1+3w)^2 \eta^2}\right)\widehat{\mu}_{\pm,k}=0,
\end{equation}
to which analytical solution in terms of the Hankel functions is known (see Appendix \ref{Appendix:spectrum} for details).\\
 
The other regime spans the time interval during which a particular mode is inside the potential, between $-\tilde{t}_c$ and $t_c$. Inside this evolution regime, we use the integral form of Eq. (\ref{wave2}) expanded in powers of $ k_{eff}$ and compute only the lowest-order term,
\begin{equation}\label{solA2}
\mu \langle \hat{Q}^{-2} \rangle^{\frac{1}{3(1-w)}} = A_1(k_{eff})+A_2(k_{eff}) \int_0^{\eta} d \eta_1 \langle \hat{Q}^{-2} \rangle^{-\frac{2}{3w-3}}.
\end{equation}
One may show that this solution exhibits the following late-time behavior:
\begin{equation}
\lim_{\tilde{t} \to +\infty} \mu \langle \hat{Q}^{-2} \rangle^{\frac{1}{3(1-w)}} = \tilde{A}_1 -\pi \tilde{A}_2 + \mathcal{O}(\tilde{t}^{n < 0}),
\end{equation}
where $\tilde{A}_1 = A_1 - \frac{\pi}{2} A_2$ and $\tilde{A}_2=A_2$.  Hence the primordial amplitude spectrum in this approximation must be proportional to this particular linear combination of constants $\tilde{A}_1$ and $\tilde{A}_2$.

The values of $\tilde{A}_1$ and $\tilde{A}_2$ are obtained from matching solutions from the two regimes at the time of the potential crossing $-\tilde{t}_c$.  On the other hand, the first regime solution is chosen by the demand that it corresponds to the Bunch-Davies vacuum in the remote past. This procedure yields $\tilde{A}_1$ subdominant and $\tilde{A}_2$ dominant as for small $\tilde{k}_{eff}$. The final result is the primordial amplitude spectrum, which is proportional to $|\tilde{A}_2|$, and reads
\begin{widetext}
\begin{equation}\label{qsp}
\delta_{\hat{h}}(\tilde{k}_{eff}) = \left(\frac{\sqrt{2 |1-3w|}}{3(1-w)} \right)^{\frac{2}{3w+1}} \Bigg| \frac{2C}{\sqrt{2 |1-3w|}} + D \Bigg| \left( \frac{\gamma}{q_b} \right)^{\frac{2w}{w-1}} \frac{1}{2} k_{max} \sqrt{\mathcal{V}_0}(1+\sigma^2)^{-\frac{1}{3w+1}}\tilde{k}_{eff}^{\frac{6w}{3w+1}},
\end{equation}
\end{widetext}
where $C$ and $D$ are constants in the general solution to Eq. (\ref{asymptotic_wave2}) inside the first evolution regime,
\begin{equation}
\begin{split}
C &= c_2 \sqrt{c_g^{\infty} k \eta_c} H^{(2)}_{\nu} (c_g^{\infty} k\eta_c), \\
D &= \frac{c_2}{2} \frac{H^{(2)}_{\nu}(c_g^{\infty} k \eta_c)}{\sqrt{c_g^{\infty} k \eta_c}} \\
 + \frac{c_2}{2} &\sqrt{c_g^{\infty} k \eta_c} \left[H^{(2)}_{\nu-1}(c_g^{\infty} k \eta_c)-H^{(2)}_{\nu+1}(c_g^{\infty} k \eta_c) \right],
\end{split}
\end{equation}
where $c_2 = \sqrt{\pi g \hbar}\ e^{-i \frac{\pi}{2}(\nu+\frac{1}{2})}$ and  $\nu = \frac{3(1-w)}{2(3w+1)}$.\\

\begin{figure}[t]
    \centering
    \includegraphics[width=0.47\textwidth]{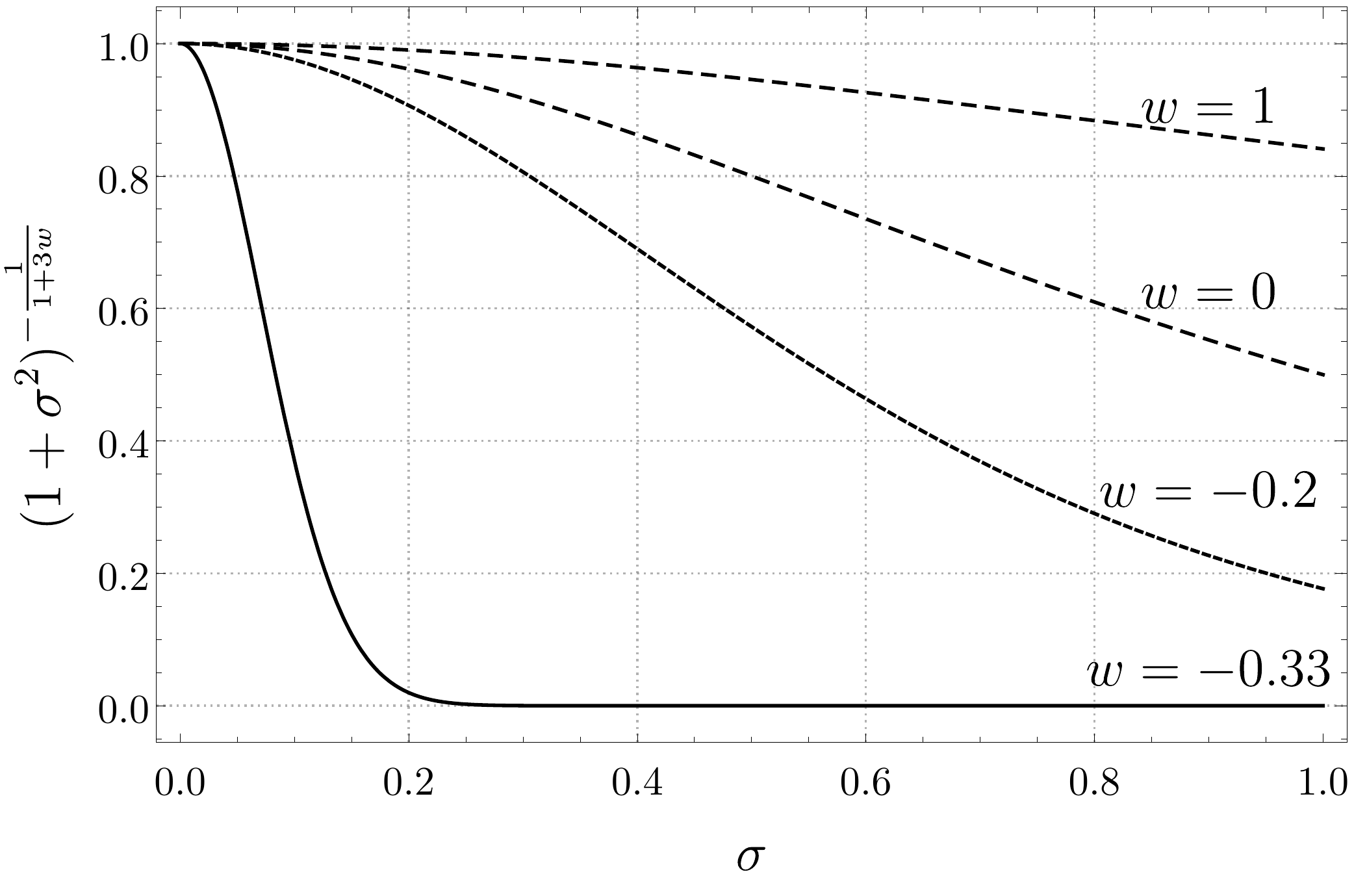}
       \caption{The suppression of primordial gravitational wave amplitude $\delta_{\hat{h}}$ in function of dispersion, $\sigma$, for a few selected fluids as given by Eq. (\ref{supp}).}
       \label{suppression}
\end{figure}

One can easily deduce from Eq. (\ref{qsp}) that the spectral index $n_t = \frac{6w}{3w+1}$ is unaffected by the spread of the background wave function. However, the absolute values of the amplitudes change under the influence of the spreading cosmological background. As $k_{eff}\eta_c$ is independent of $\sigma$, the relation between the semiclassical and the ``quantum" amplitude spectrum reads
\begin{equation}\label{supp}
\delta_{\hat{h}}(\tilde{k}_{eff}) = (1+\sigma^2)^{-\frac{1}{3w+1}} \cdot \delta_{\hat{h}}(\tilde{k}_{eff}) \Big|_{\sigma=0}.
\end{equation}
The quantum factor in the above equation takes values from the interval $0 < (1+\sigma^2)^{\frac{3w}{3w+1}} \leq 1$ for fluids with $-\frac{1}{3} \leq w \leq 1$. In Fig. \ref{suppression} we plot the dependence of the gravity-wave amplitude on the dispersion $\sigma$ for selected values of $w$. It universally leads to the quantum dampening of the amplitude. The suppression may be mild and rather irrelevant or, in the case when $w\approx -\frac{1}{3}$, large and significant leading to observable effects.

{ We found that the quantum spread does not affect the spectral index. We expect this to be a quite general property that follows from the fact that the  long-wavelength modes enter the potential when the potential is still classical. Thus, the constants in Eq. (\ref{solA2}), $A_1$ and in particular $A_2$ that becomes dominant for small ${k}_{eff}$, are determined already in the classical universe. On the other hand, the integral in Eq. (\ref{solA2}) does not depend on ${k}_{eff}$ but may be sensitive to the details of the dynamics in the vicinity of the bounce. Hence, the quantum spread can alter the amplitude of the long-wavelength modes by an overall factor that does not depend on scale.}

\section{Conclusions}
We studied the effect of the bounce and quantum uncertainties in the background geometry on the gravitational waves propagating across the primordial universe. We first analyzed the semiclassical description in which the wave packets are assumed to be infinitely narrow. We reproduced a class of interaction potentials for gravitational waves that had been previously obtained within the Bohm-de Broglie trajectory approach. We obtained the amplitude spectra in agreement with the results of \cite{Peter2006}. Next we studied a free parameter of our model $K$ and the way it determines the redshift of the bounce and the matter density at the bounce given the known bound on the primordial gravitational waves amplitude at the physically relevant scales. We found a large space of admissible parameters for all cosmological fluids, which produced plausible cosmological scenarios.

We then enhanced the treatment by the inclusion of quantum spread. By employing the WKB approximation we were able to obtain an analytical relation between dispersion, $\sigma$, and the evolution of gravitational waves. We found that the spread  induces qualitative changes to the dynamical law of gravitational waves. One way the uncertainties enter the dynamical law is by varying the speed of gravitational waves as the universe bounces. The other way they manifest themselves is by altering the interaction potential. We found that these important changes of the dynamical law ultimately have no effect on the spectral index of the primordial amplitude for long-wavelength modes. The amplitude on the other hand becomes multiplied by an overall factor independent of the wavelength. The factor, however, is rather irrelevant for most cosmological fluids; nevertheless, it can cause a significant suppression of the amplitude in cases when $w\approx -\frac{1}{3}$.\\

The finding that the quantum spread does not influence the cosmological predictions for most cosmological fluids is very important theoretically. It implies that the semiclassical analysis is completely sufficient in those cases at linear order. It might also imply that it will never be possible to discern any difference between classical and quantum bounce scenarios.

The finding that the quantum spread may significantly suppress the primordial amplitude for some cosmological fluids, even if they themselves are not physically appealing, indicates that one should verify the possible effect of quantum spread every time that one introduces a form of matter not included in our work. We note that it is not clear from our work why the amplitude is suppressed rather than amplified. It might be the case that there exist quantum states of the background spacetime that amplify the amplitude of gravitational waves.

Finally, let us observe that the effect of uncertainties illustrates the basic fact about quantum mechanics and semiclassical descriptions thereof. Namely, there are infinitely many ways in which one can replace a given classical observable with a function of expectation values of operators that behaves like the classical observable for large universes. All such functions provide semiclassical expressions for a given classical observable but with a different behavior exactly in the regime where classical mechanics breaks down. This is illustrated by the examples of the scale factor described in the Introduction and shows a serious limitation on the physical interpretation of semiclassical descriptions. It is also illustrated by the nontrivial evolution of the coupling of the gravitational waves and indicates that any semiclassical description must be verified whether it indeed reproduces the correct quantum behavior.

\begin{acknowledgments}
The work of P.M. is part of the research project 2018/30/E/ST2/00370 financed by National Science Centre (NCN), Poland. A.M. acknowledges the financing from the research project 2017/27/N/ST2/01964 by National Science Centre (NCN), Poland. The authors thank Herv\'e Bergeron and Patrick Peter for helpful discussions.
\end{acknowledgments}

\vspace{.1cm}

\appendix
\section{GRAVITY-WAVE PROPAGATION EQUATION IN WKB APPROXIMATION}\label{wkbapp}
In the WKB approximation the gravity-wave propagation equation (\ref{wave2}) reads
\begin{align}
\widehat{\mu}_{\pm,\vec{k}}''+\left(k^2c^2_g(t(\eta))-V(t(\eta))\right)\widehat{\mu}_{\pm,\vec{k}}=0,
\end{align}
where $\eta(t)=\int^t\left(\frac{\gamma^2}{8q_b^2\sigma}\ln\bigg|\frac{(1+\sigma)^4+(k_{max}t')^2}{(1-\sigma)^4+(k_{max}t')^2}\bigg|\right)^{\frac{3w-1}{3w-3}}\ud t'$ and
\begin{widetext}
\begin{align}\begin{split}
V(t) &= \frac{-256  (k_{max}t)^2 (\sigma + \sigma^3)^2 (-5 + 6 w)(8 \sigma)^{\frac{2 - 6 w}{
 3 - 3 w}}\Lambda^{\frac{12 w - 8}{3 - 3 w}}k_{max}^2\big(\frac{q_b^2}{\gamma^2}\big)^{\frac{6w-2}{3w-3}}}{ ((k_{max}t)^2 + (1-\sigma)^4)^2 ((k_{max}t)^2 + (1 + \sigma)^4)^2 (3w-3)^2} \\
  &+ \frac{16 \sigma (1 + \sigma^2) ((1-\sigma^2)^4 - 2 (k_{max}t)^2 (1 + 6 \sigma^2 + \sigma^4)-3 (k_{max}t)^4 )(8 \sigma)^{\frac{2 - 6 w}{
 3 - 3 w}} \Lambda^{\frac{9 w - 5}{3 - 3 w}}k_{max}^2\big(\frac{q_b^2}{\gamma^2}\big)^{\frac{6w-2}{3w-3}}}{((k_{max}t)^2 + (1-\sigma)^4)^2 ((k_{max}t)^2 + (1 +  \sigma)^4)^2 (3 - 3 w)},\end{split}
\end{align} 
where $\Lambda=\ln\bigg|\frac{
      (k_{max}t)^2 + (\sigma+1)^4}{(k_{max}t)^2 + ( \sigma-1)^4}\bigg|$,
\begin{align}\begin{split}
c_g^2(t)=\frac{ 4^{\frac{1+w}{w-1}}(3-3 w) \Gamma \left(\frac{2}{3-3 w}\right)\left(F_{-}-F_{+}\right) \bigg(\frac{\ln \big|\frac{(1+\sigma )^4+(k_{max}t)^2}{(\sigma -1)^4+(k_{max}t)^2}\big|}{\sigma }\bigg)^{\frac{3 w+1}{3-3 w}}}{\left(1-\sigma ^2\right)^{\frac{4}{3-3 w}} \left(1+\sigma ^2\right)^{\frac{3 w+1}{3-3 w}} \left((1-\sigma )^{\frac{4 w}{w-1}} (1+\sigma)^{\frac{4}{3 (w-1)}}-(1-\sigma )^{\frac{4}{3 (w-1)}} (1+\sigma)^{\frac{4 w}{w-1}}\right) \Gamma \left(\frac{1-3 w}{3 (w-1)}\right)},
\end{split}
\end{align}
where $F_{\pm}=(1\pm\sigma )^{\frac{8}{3-3 w}}~_2F_1\left(\frac{3 w+1}{3 w-3},\frac{2}{3w-3} ;\frac{1-3w}{3-3w} ;-\frac{(k_{max}t)^2}{(\sigma \pm 1)^4}\right)$.
\end{widetext}

\section{DERIVATION OF THE GRAVITY-WAVE AMPLITUDE }\label{Appendix:spectrum}
We follow the notation introduced in Sec.  \ref{lwa}. In the first evolution regime the potential is subdominant and the time parameter satisfies $|\tilde{t}| \geq \tilde{t}_c$. The asymptotic behavior for the expectation value of the inverse position squared is
\begin{equation}
\lim_{\tilde{t} \to \pm \infty} \langle \hat{Q}^{-2} \rangle = \frac{1+ \sigma^2}{q_b^2 \tilde{t}^2},
\end{equation}
which, by the virtue of Eq. (\ref{eta_t}), yields the asymptotic relation between the time parameters $\tilde{t}$ and $\eta$,
\begin{equation}
\begin{split}
\lim_{\tilde{t} \to \pm \infty} k_{max} \eta(\tilde{t}) &= 
\\= \pm \frac{3(1-w)}{3w+1} &|\tilde{t}|^{\frac{1+3w}{3(1-w)}} \left(1+\sigma^2\right)^{\frac{3w-1}{3(w-1)}}  \left(\frac{\gamma^2 }{q_b^2}\right)^{\frac{3w-1}{3(w-1)}}.
\end{split}
\end{equation}
Hence, the asymptotic form of Eq. (\ref{wave2}) reads
\begin{equation}
\widehat{\mu}_{\pm,k}''+\left(\left(c_g^{\infty}k\right)^2+ \frac{2(3w-1)}{(1+3w)^2 \eta^2}\right)\widehat{\mu}_{\pm,k}=0,
\end{equation}
the solution of which is
\begin{equation}\label{solution_Hankel}
\mu = \sqrt{\eta} \left[ c_1(k) H_{\nu}^{(1)}(c_g^{\infty} k\eta)+ c_2(k) H_{\nu}^{(2)}(c_g^{\infty} k\eta) \right],
\end{equation}
where $\nu = \frac{3(1-w)}{2(3w+1)}$. The solution (\ref{solution_Hankel}) is a sufficiently accurate solution of Eq. (\ref{wave2}) for $|\tilde{t}| \gg 1$ and is used  at all times for which the potential is subdominant. This happens before (or after) the potential crossing time $-\tilde{t}_c$ ($\tilde{t}_c$), which for small $\tilde{k}_{eff}$ modes reads
\begin{equation}
\tilde{t}_c = \left[\frac{9}{2} \frac{(1-w)^2}{|1-3w|} \tilde{k}^2_{eff} \right]^{\frac{3(w-1)}{2(1+3w)}} \left[1+\sigma^2 \right]^{\frac{3w-1}{3w+1}},
\end{equation}
where $\tilde{k}_{eff} = \tilde{k} c_g^{\infty}$.\\
On the other hand, at times $-\tilde{t}_c \leq \tilde{t} \leq \tilde{t}_c$ when the interaction potential is dominant, the solution is approximated by the lowest order in $k_{eff}$ terms of the formal solution of Eq. (\ref{wave2}),
\begin{equation}\label{intsol}
\begin{split}
\mu_k \langle \hat{Q}^{-2} \rangle^{\frac{1}{3(1-w)}} = A_1(k_{eff})+A_2(k_{eff}) \int d \eta_1 \langle \hat{Q}^{-2} \rangle^{-\frac{2}{3w-3}}\\ - k_{eff}^2 \int d\eta_1 \langle \hat{Q}^{-2} \rangle^{-\frac{2}{3w-3}} \int d\eta_2 \langle \hat{Q}^{-2} \rangle^{\frac{1}{3w-3}} \mu_k ,
\end{split}
\end{equation}
where the quantity  $\mu_k \langle \hat{Q}^{-2} \rangle^{\frac{1}{3(1-w)}} \gamma^{\frac{2}{3(1-w)}}$ corresponds to the classical variable $h_k=\frac{\mu_k}{a}$. Because the change in $c_g^2$ does not break the dominance of the interaction potential, the present approximation neglects the evolution of $c_g^2$ and picks its value at infinity where the effective wave number $\tilde{k}_{eff} = \tilde{k} c_g^{\infty}$ is defined. The asymptotic value of speed of gravitational waves [see the definition below Eq. (\ref{wave2})],
\begin{equation}
c_g^{\infty} = \langle \hat{Q}^{\frac{6w+2}{3(1-w)}} \rangle_{\infty} \langle \hat{Q}^{-2} \rangle_{\infty}^{\frac{3w+1}{3(1-w)}},
\end{equation} 
is easily found in the WKB approximation. Indeed, as one may show,
\begin{equation}
\langle \hat{Q}^n \rangle_{\infty} = \frac{\tilde{t}^n q_b^n}{2(2-n)} \left[(1+\sigma)^{2-n}-(1-\sigma)^{2-n} \right].
\end{equation}

After neglecting higher-order terms in Eq. (\ref{intsol}), the only integral left reads
\begin{equation}
\begin{split}
\frac{8\sigma q_b^2}{k_{max}} \int d \eta_1 \langle \hat{Q}^{-2} \rangle^{-\frac{2}{3w-3}} = 2(1+\sigma)^2 \arctan\left[ \frac{\tilde{t}}{(1+\sigma)^2} \right] \\
-2(1-\sigma)^2 \arctan\left[ \frac{\tilde{t}}{(1-\sigma)^2} \right]
+\tilde{t} \ln\left[ \frac{(1+\sigma)^4 + \tilde{t}^2}{(1-\sigma)^4 + \tilde{t}^2} \right].
\end{split}
\end{equation}
Therefore, the solution far away from the bounce, in the leading terms, is
\begin{equation}\label{asymptotic_va}
\begin{split}
\lim_{\tilde{t} \to -\infty} \mu_k \langle \hat{Q}^{-2} \rangle^{\frac{1}{3(1-w)}} &= \tilde{A}_1 + \tilde{A}_2 \frac{1+\sigma^2}{\tilde{t}},\\
\lim_{\tilde{t} \to +\infty} \mu_k \langle \hat{Q}^{-2} \rangle^{\frac{1}{3(1-w)}} &= \tilde{A}_1 -\pi \tilde{A}_2 + \mathcal{O}(\tilde{t}^{n < 0}).
\end{split}
\end{equation}
We match the solutions at the point $-\tilde{t}_c$, where $\mu_k$ can be approximated by
\begin{equation}\label{matching_1}
\begin{split}
\mu_k(-\tilde{t}_c) = \tilde{A}_1 (-\tilde{t}_c)^{\frac{2}{3(1-w)}} \left(\frac{1+\sigma^2}{q_b^2} \right)^{\frac{1}{3(w-1)}} \\
 + \tilde{A}_2 (-\tilde{t}_c)^{\frac{3w-1}{3(1-w)}} \left(\frac{1+\sigma^2}{q_b^2} \right)^{\frac{3w-2}{3(w-1)}}
\end{split}
\end{equation}
and, from (\ref{solution_Hankel}), we also have [$\eta_c = \eta (\tilde{t}_c)$]
\begin{equation}\label{matching_2}
\mu_k(-\eta_c) = \frac{C}{\sqrt{c_g^{\infty} k}},\ \mu_k'(-\eta_c) = D \sqrt{c_g^{\infty} k}
\end{equation}
Assuming the Bunch-Davies vacuum normalization $c_1=0$ and $c_2 = \sqrt{\pi g \hbar}\ e^{-i \frac{\pi}{2}(\nu+\frac{1}{2})}$, the constants are
\begin{equation}
\begin{split}
C &= c_2 \sqrt{c_g^{\infty} k \eta_c} H^{(2)}_{\nu} (c_g^{\infty} k\eta_c), \\
D &= \frac{c_2}{2} \frac{H^{(2)}_{\nu}(c_g^{\infty} k \eta_c)}{\sqrt{c_g^{\infty} k \eta_c}} \\
 + \frac{c_2}{2} &\sqrt{c_g^{\infty} k \eta_c} \left[H^{(2)}_{\nu-1}(c_g^{\infty} k \eta_c)-H^{(2)}_{\nu+1}(c_g^{\infty} k \eta_c) \right].
\end{split}
\end{equation}
Combining Eqs (\ref{matching_1}) and (\ref{matching_2}) we obtain
\begin{equation}
\begin{split}
\tilde{A}_1 &= -\left(-\frac{\sqrt{2 |1-3w|}}{3(1-w)} \right)^{\frac{3w-1}{3w+1}} (1+\sigma^2)^{\frac{1}{3w+1}}  \\
\times & \frac{\gamma^{\frac{1-3w}{3(1-w)}}}{q_b \sqrt{k_{max}}} \left( \frac{(1-3w)C}{\sqrt{2|1-3w|}}-D \right) \tilde{k}_{eff}^{\frac{3(1-w)}{2(3w+1)}}, \\
\tilde{A}_2 &= \left(-\frac{\sqrt{2 |1-3w|}}{3(1-w)} \right)^{\frac{2}{3w+1}} (1+\sigma^2)^{-\frac{1}{3w+1}}  \\
\times & \frac{\gamma^{\frac{1-3w}{3(1-w)}}}{q_b \sqrt{k_{max}}} \left( \frac{-2C}{\sqrt{2|1-3w|}}-D \right) \tilde{k}_{eff}^{\frac{3(w-1)}{2(3w+1)}}. \\
\end{split}
\end{equation}
Note that for cosmological fluids with $-\frac{1}{3} \leq w \leq 1$ the coefficient $\tilde{A}_1$ scales with a positive power of $\tilde{k}$; therefore, it is subdominant for $\tilde{k} \ll 1$. The dominant part of the amplitude spectrum is determined by $\tilde{A}_2$. Making use of  Eq. (\ref{asymptotic_va}), we find the spectrum of amplitude (\ref{spdef}),
\vspace{0.1 cm}
\begin{widetext}
\begin{equation}
\begin{split}
\delta_{\hat{h}}(\tilde{k}_{eff}) = \left(\frac{\sqrt{2 |1-3w|}}{3(1-w)} \right)^{\frac{2}{3w+1}} \Bigg| \frac{2C}{\sqrt{2 |1-3w|}} + D \Bigg| \left( \frac{\gamma}{q_b} \right)^{\frac{2w}{w-1}} \frac{1}{2} k_{max} \sqrt{\mathcal{V}_0}(1+\sigma^2)^{-\frac{1}{3w+1}}\tilde{k}_{eff}^{\frac{6w}{3w+1}}.
\end{split}
\end{equation}
\end{widetext}

\section{ASYMPTOTIC EXPANSION OF THE AMPLITUDE SPECTRUM}\label{adiabatic}
{ In the present section we apply the adiabatic subtraction described in Chap. 3 of \cite{parker} to regularize the amplitude spectrum (\ref{spdef}) based on the adiabatic expansion of the mode functions ${\mu}_{\pm,\vec{k}}$ that satisfy Eq.  (\ref{wave2}). The obtained result equally applies to the semiclassical model with $c_g=1$. The amplitude spectrum is obtained from the power spectrum that is quadratic in the modes ${\mu}_{k}$. Given that the modes satisfy the initial condition (\ref{ini}) for $t_0\rightarrow -\infty$, we obtain
\begin{align}
|{\mu}_{k}|^2=\frac{4\mathfrak{g}\hbar}{W},
\end{align}
where $W^{-1} = (W^{-1})^{(0)} + (W^{-1})^{(2)} + O(T^{-4})$
with the index $^{(n)}$ denoting the order of the adiabatic expansion. We find
\begin{align}\begin{split}
(W^{-1})^{(0)} &=  c_g^{-1} k^{-1},\\
(W^{-1})^{(2)} &= \frac{c_g^{-3}}{2} \left(\frac{\left( \langle \hat{Q}^{-2} \rangle^{\frac{1}{3w-3}} \right)''}{\langle \hat{Q}^{-2} \rangle^{\frac{1}{3w-3}}} -c_g^{\frac{1}{2}} (c_g^{-\frac{1}{2}})'' \right)k^{-3},\end{split}
\end{align}
where $c_g$ is defined below Eq. (\ref{wave2}). If we were to compute the two-point correlation function $\langle 0|\widehat{\mu}(x)\widehat{\mu}(x')|0\rangle\propto\int \ud^3 k \frac{e^{i\vec{k}(x-x')}}{W}$ the zero- and second-order terms  $(W^{-1})^{(0)}$ and $(W^{-1})^{(2)}$ would produce, respectively, quadratic and logarithmic divergences as $x\rightarrow x'$. The next-order term $(W^{-1})^{(4)}$ scales at least as $k^{-5}$; therefore, it must give a finite contribution.

Finally, the regularized amplitude spectrum is found to read [cf. Eq. (\ref{spdef})],
\begin{equation}
\delta_{\widehat{h}}(k)=\frac{\sqrt{|\mu_{k}|^2-4\mathfrak{g}\hbar(W^{-1})^{(0)} - 4\mathfrak{g}\hbar(W^{-1})^{(2)}}}{2\pi \mathcal{V}_0^{-\frac{1}{2}} \langle\left(\frac{\widehat{Q}}{\gamma}\right)^{-2}\rangle^{\frac{1}{3w-3}}}k^{\frac{3}{2}}~.
\end{equation}
}

\end{document}